\documentclass[rog]{agutex}

\usepackage{amsmath}
\usepackage{amssymb}
\newcommand{\about} {\mathop{\sim}\!}
\newcommand{\Ro}{\mathrm{Ro}}
\renewcommand{\vec}[1]{\boldsymbol{\mathrm{#1}}}   

\usepackage{agu-ps}

  \usepackage[pdftex]{graphicx}
  \graphicspath{{figures/}}

 \setkeys{Gin}{draft=false}
\authorrunninghead{SCHNEIDER ET AL.}

\titlerunninghead{WATER VAPOR AND CLIMATE CHANGE}

\authoraddr{Tapio Schneider, %
  California Institute of Technology, Pasadena, CA 91125-2300, USA.
  (tapio@caltech.edu)}

\authoraddr{Paul A. O'Gorman, %
  Massachusetts Institute of Technology, Cambridge, MA 02139, USA.
  (pog@mit.edu)}

\authoraddr{Xavier J. Levine, %
  California Institute of Technology, Pasadena, CA 91125-2300, USA.
  (xavier@caltech.edu)}
\begin{document}

%
%

\title{Water vapor and the dynamics of climate changes}
%

%
%


\author{Tapio Schneider}
\affil{California Institute of Technology, Pasadena, California, USA}

\author{Paul A. O'Gorman}
\affil{Massachusetts Institute of Technology, Cambridge, Massachusetts, USA}

\author{Xavier Levine}
\affil{California Institute of Technology, Pasadena, California, USA}

%
%


\begin{abstract}
  Water vapor is not only Earth's dominant greenhouse gas. Through the
  release of latent heat when it condenses, it also plays an active
  role in dynamic processes that shape the global circulation of the
  atmosphere and thus climate. Here we present an overview of how
  latent heat release affects atmosphere dynamics in a broad range of
  climates, ranging from extremely cold to extremely warm. Contrary to
  widely held beliefs, atmospheric circulation statistics can change
  non-monotonically with global-mean surface temperature, in part
  because of dynamic effects of water vapor. For example, the
  strengths of the tropical Hadley circulation and of zonally
  asymmetric tropical circulations, as well as the kinetic energy of
  extratropical baroclinic eddies, can be lower than they presently
  are both in much warmer climates and in much colder climates. We
  discuss how latent heat release is implicated in such circulation
  changes, particularly through its effect on the atmospheric static
  stability, and we illustrate the circulation changes through
  simulations with an idealized general circulation model. This allows
  us to explore a continuum of climates, constrain macroscopic laws
  governing this climatic continuum, and place past and possible
  future climate changes in a broader context.
\end{abstract}

%
%

%

\begin{article}

%
%

\section{Introduction}

Water vapor is not only important for Earth's radiative balance as the
dominant greenhouse gas of the atmosphere. It is also an active player
in dynamic processes that shape the global circulation of the
atmosphere and thus climate. The latent heat released when atmospheric
water vapor condenses and the cooling of air through evaporation or
sublimation of condensate affect atmospheric circulations. Although
the mechanisms are not well understood, it is widely appreciated that
heating and cooling of air through phase changes of water are integral
to moist convection and dynamics in the equatorial region. But that
water vapor plays an active and important role in dynamics globally is
less widely appreciated, and how it does so is only beginning to be
investigated. For instance, there is evidence that the width of the
Hadley circulation has increased over the past decades
\citep[e.g.,][]{Hu07b,Seidel07,Seidel08}, and it also increases in
many simulations of climate change in response to increased
concentrations of greenhouse gases
\citep[e.g.,][]{Kushner01,Lu07,Previdi07,Johanson09}. This widening of
the Hadley circulation is often linked to the decrease in the
moist-adiabatic temperature lapse rate with increasing surface
temperature, which results in an increased tropical static stability
and can lead to a widening of the Hadley circulation, at least in dry
atmospheres \citep[e.g.,][]{Held00b,Walker06,Frierson07d,Korty08}. Yet
it is unclear how the width of the Hadley circulation in an atmosphere
in which water vapor is dynamically active relates to the static
stability, or in fact, how the static stability thought to be
relevant---that at the subtropical termini of the Hadley
circulation---is controlled.

Here we present an overview of dynamic effects of water vapor in the
global circulation of the atmosphere and in climate changes. What may
be called water vapor kinematics---the study of the distribution of
water vapor given the motions of the atmosphere---has recently been
reviewed by \citet{Held00a}, \citet{Pierrehumbert07}, and
\citet{Sherwood09}. We bracket off questions of water vapor kinematics
to the extent possible and instead focus on what may be called water
vapor dynamics---the study of the dynamic effects of heating and
cooling of air through phase changes of water.

Our emphasis lies on large scales, from the scales of extratropical
storms ($\about 1000\,\mathrm{km}$) to the planetary scale of the
Hadley circulation. In motions on such large scales, the release of
latent heat through condensation generally is more important than the
cooling of air through evaporation or sublimation of condensate: The
residence times of vapor and condensate are similar (days and longer),
and so are the specific latent heat of vaporization and that of
sublimation, but the atmosphere in the global mean contains about 250
times more water vapor ($\about 25\,\mathrm{kg\,m^{-2}}$) than liquid
water and ice ($\about 0.1\,\mathrm{kg\,m^{-2}}$)
\citep{Trenberth05a}. Nonetheless, even motions on large scales are
affected by smaller-scale dynamics such as moist convection, for which
cooling through evaporation of condensate and the convective
downdrafts thereby induced are essential
\citep[e.g.,][]{Emanuel94b}. The emphasis on large scales allows us to
sideline some of the complexities of moist convection and consider
only the collective effect of many convective cells on their
large-scale environment, assuming that the convective cells adjust
rapidly to their environment and so are in statistical equilibrium
(``quasi-equilibrium'') with it \citep{Arakawa74}. Our reasoning about
the effect of moist convection on large-scale motions builds upon the
cornerstone of convective quasi-equilibrium dynamics, well supported
by observations and simulations of radiative-convective equilibrium:
convection, where it occurs, tends to establish a thermal
stratification with moist-adiabatic temperature lapse rates (see
\citet{Emanuel94b}, \citet{Emanuel07}, and \citet{Neelin08} for
overviews).

Dynamic effects of water vapor in the global circulation of the
atmosphere have typically been discussed in the context of specific
past climates, such as that of the Last Glacial Maximum (LGM), or
possible future climate changes in response to increased
concentrations of greenhouse gases. We view past and possible future
climates as parts of a climatic continuum that is governed by
universal, albeit largely unknown, macroscopic laws. Our goal is to
constrain the forms such macroscopic laws may take. They cannot be
inferred from observational data, as it can be misleading to infer
laws governing climate changes from fluctuations within the present
climate (e.g., from El Ni{\~n}o and the Southern Oscillation, as we
will discuss further below). And they are difficult to infer from
simulations with comprehensive climate models, whose complexity can
obscure the chain of causes and effects in climate changes.

Therefore, we illustrate theoretical developments in what follows with
simulations of a broad range of climates with an idealized general
circulation model (GCM). The simulations are described in detail in
\citet{OGorman08b}. They are made with a GCM similar to that of
\citet{Frierson06a}, containing idealized representations of dynamic
effects of water vapor but not accounting for complexities not
directly related to water vapor dynamics. For example, the GCM has a
surface that is uniform and water covered (a ``slab ocean'' that does
not transport heat), and there are no topography and no radiative
water vapor or cloud feedbacks. The GCM employs a variant of the
quasi-equilibrium moist convection scheme of \citet{Frierson07b}, has
insolation fixed at perpetual equinox, and takes only the vapor-liquid
phase transition of water into account, assuming a constant specific
latent heat of vaporization. Consistently but unlike what would occur
in the real world, ice is ignored in the model, be it as cloud ice,
sea ice, or land ice. We obtained a broad range of statistically
steady, axisymmetric, and hemispherically symmetric climates by
varying the optical thickness of an idealized atmospheric longwave
absorber, keeping shortwave absorption fixed and assuming gray
radiative transfer. The climates have global-mean surface temperatures
ranging from 259~K (pole-equator surface temperature contrast 70~K) to
316~K (temperature contrast 24~K) and atmospheric water vapor
concentrations varying by almost two orders of magnitude. We will
discuss dynamic effects of water vapor in past and possible future
climates in the context of this broad sample from a climatic
continuum, making connections to observations and more comprehensive
GCMs wherever possible. This allows us to examine critically, and
ultimately to reject, some widely held beliefs, such as that the
Hadley circulation would generally become weaker as the climate warms,
or that extratropical storms would generally be stronger than they are
today in a climate like that of the LGM with larger pole-equator
surface temperature contrasts.

Section~\ref{s:constraints} reviews energetic constraints on the
concentration of atmospheric water vapor and precipitation as
background for the discussion of how water vapor dynamics affects
climate changes.  Section~\ref{s:tropics} examines tropical
circulations, with emphasis on the Hadley
circulation. Section~\ref{s:extratropics} examines extratropical
circulations, with emphasis on extratropical storms and the static
stability, which occupies a central place if one wants to understand
the effects of water vapor on extratropical dynamics.
Section~\ref{s:summary} summarizes conclusions and open questions.

\section{Energetic constraints on water vapor concentration and
  precipitation}\label{s:constraints}

Water vapor dynamics is more important in warmer than in colder
climates because the atmospheric water vapor concentration generally
increases with surface temperature. This is a consequence of the rapid
increase of the saturation vapor pressure with temperature. According
to the Clausius-Clapeyron relation, a small change $\delta T$ in
temperature $T$ leads to a fractional change $\delta e^*/e^*$ in
saturation vapor pressure $e^*$ of
\begin{equation}\label{e:cc}
  \frac{\delta e^*}{e^*} \approx \frac{L}{R_v T^2} \, \delta T,
\end{equation}
where $R_v$ is the gas constant of water vapor and $L$ is the specific
latent heat of vaporization. If one substitutes temperatures
representative of near-surface air in the present climate, the
fractional increase in saturation vapor pressure with temperature is
about $6$--$7\% \, \mathrm{K^{-1}}$, that is, the saturation vapor
pressure increases $6$--$7\%$ if the temperature increases
$1\,\mathrm{K}$ (e.g., \cite{Boer93,Wentz00,Held00a,Trenberth03b}). In
Earth's atmosphere in the past decades, precipitable water
(column-integrated specific humidity) has varied with surface
temperature at a rate of $7$--$9\% \, \mathrm{K^{-1}}$, averaged over
the tropics or over all oceans \citep{Wentz00,Trenberth05b}. Thus, the
fractional variations in precipitable water are similar to those in
near-surface saturation vapor pressure. They are consistent with an
approximately constant effective relative humidity---the ratio of
column-integrated vapor pressure to saturation vapor pressure, or the
relative-humidity average weighted by the saturation vapor pressure,
i.e., weighted toward the lower troposphere. Similarly, in simulations
of climate change scenarios, global-mean precipitable water increases
with global-mean surface temperature at a rate of $\about 7.5\%\,
\mathrm{K^{-1}}$, likewise consistent with an approximately constant
effective relative humidity \citep{Held06,Willett07,Stephens08a}.

But global-mean precipitation and evaporation (which are equal in a
statistically steady state) increase more slowly with temperature than
does precipitable water. In simulations of climate change scenarios,
global-mean precipitation and evaporation increase with global-mean
surface temperature at a rate of only
$2$--$3\%\,\mathrm{K^{-1}}$---considerably less than the rate at which
precipitable water increases
\citep[e.g.,][]{Knutson95,Allen02,Held06,Stephens08a}.  They have
varied with surface temperature at a similar rates in Earth's
atmosphere in the past decades \citep{Adler08}. This points to
energetic constraints on the global-mean precipitation and evaporation
\citep{Boer93}.

The surface energy balance closely links changes in evaporation to
changes in near-surface saturation specific humidity and relative
humidity. The evaporation $E$ enters the surface energy balance as the
latent heat flux $LE$, which, in Earth's present climate, is the
largest loss term balancing the energy gained at the surface through
absorption of solar radiation \citep{Kiehl97,Trenberth09a}. The
evaporation is related to the specific humidity $q$ near the surface
and the saturation specific humidity $q_s^*$ at the surface by the
bulk aerodynamic formula,
\begin{equation}
  E \approx \rho C_W \|\vec{v} \| (q_s^* - q).
\end{equation}
Here, $\rho$ is the density of near-surface air, $\vec{v}$ is the
near-surface wind, $C_W$ is a bulk transfer coefficient, and the
formula is valid over oceans, where most evaporation occurs (e.g.,
\cite{Peixoto92}). Over oceans, the disequilibrium factor $q_s^* - q$
between the surface and near-surface air is usually dominated by the
subsaturation of near-surface air, rather than by the temperature
difference between the surface and near-surface air; therefore, it can
be approximated as $q_s^* - q \approx (1-\mathcal{H}) q_s^*$, with
near-surface relative humidity $\mathcal{H}$. Changes in near-surface
relative humidity $\delta\mathcal{H}$ can then be related to
fractional changes in evaporation $\delta E/E$ and near-surface
saturation specific humidity $\delta q_s^*/q_s^*$ if we make two
simplifying assumptions: (i) changes in evaporation with climate are
dominated by changes in the disequilibrium factor $q_s^* - q$, and
(ii) changes in the disequilibrium factor $q_s^* - q$, in turn, are
dominated by changes in near-surface relative humidity and saturation
specific humidity, so that $\delta(q_s^* - q) \approx (1-\mathcal{H})
\delta q_s^* - q_s^* \delta\mathcal{H}$.
This leads to
\begin{equation}\label{e:rh_change}
  \delta\mathcal{H} \approx (1 - \mathcal{H})\left(\frac{\delta
      q_s^*}{q_s^*} - \frac{\delta E}{E}\right),
\end{equation}
an expression equivalent to one used by \cite{Boer93} to evaluate
hydrologic-cycle changes in climate change simulations.

As discussed by \cite{Boer93} and \cite{Held00a}, the relation
\eqref{e:rh_change} together with the surface energy balance constrain
the changes in evaporation and near-surface relative humidity that are
possible for a given change in radiative forcing and
temperature. Assume evaporation increases with surface temperature at
$2.5\%\,\mathrm{K^{-1}}$ in the global mean, and saturation vapor
pressure increases at $6.5\%\,\mathrm{K^{-1}}$, as it does in typical
climate change simulations. Then, if the global-mean surface
temperature increases by $3\,\mathrm{K}$, the global-mean evaporation
increases by $\delta E/E \approx 7.5\%$, and the saturation specific
humidity at the surface increases by $\delta q_s^*/q_s^* \approx
\delta e^*/e^* \approx 19.5\%$.  To the extent that the relation
\eqref{e:rh_change} is adequate and for a near-surface relative
humidity of $80\%$, it follows that the relative humidity
$\mathcal{H}$ increases by about $\delta
\mathcal{H}=(1-0.8)(19.5-7.5)=2.4$ percentage points---a comparatively
small change. The precise magnitude of the relative humidity changes
depends on changes in the surface winds and in the temperature
difference between the surface and near-surface air. However, even if,
for example, changes in the temperature difference between the surface
and near-surface air influence the disequilibrium factor $q_s^* - q$
similarly strongly as changes in the near-surface relative humidity,
the order of magnitude of the terms shows that the near-surface
relative humidity generally changes less than the near-surface
saturation specific humidity. This is especially the case if the
near-surface air is close to saturation, so that the factor
$(1-\mathcal{H})$ in \eqref{e:rh_change} is small.

Because most water vapor in the atmosphere is confined near the
surface (the water vapor scale height is $\about 2\,\mathrm{km}$), the
fact that changes in near-surface relative humidity are constrained to
be relatively small implies that changes in precipitable water are
dominated by changes in the near-surface saturation specific humidity.
Hence, precipitable water changes scale approximately with the rate
given by the Clausius-Clapeyron relation \eqref{e:cc}, as seen in
observed climate variations and simulated climate change
scenarios. Free-tropospheric relative humidity need not stay fixed,
however, so precipitable water changes may deviate slightly from
Clausius-Clapeyron scaling.

It is also clear that the rate of change of evaporation with
global-mean surface temperature cannot differ vastly from the
$2$--$3\%\,\mathrm{K^{-1}}$ quoted above, as would be necessary for
significant relative humidity changes. To illustrate how strongly
changes in evaporation and near-surface relative humidity are
constrained by the surface energy balance, consider a hypothetical
case that will turn out to be impossible: assume an increase in the
concentration of greenhouse gases would lead to a 3-K global-mean
surface temperature increase in a statistically steady state,
accompanied by a global-mean saturation specific humidity increase at
the surface by $\about 19.5\%$; assume further that this would lead to
a reduction in near-surface relative humidity from $80\%$ to
$70\%$. According to \eqref{e:rh_change}, evaporation would then have
to increase by $\about 70\%$ in the global mean. Currently total
evaporation at Earth's surface amounts to a latent heat flux of about
$80\,\mathrm{W\, m^{-2}}$ \citep{Kiehl97,Trenberth09a}. A $70\%$
increase would imply that an additional energy flux of
$56\,\mathrm{W\,m^{-2}}$ would have to be available to the surface to
balance the additional evaporation. The global-mean net irradiance
would have to increase and/or the upward sensible heat flux at the
surface would have to decrease by this amount. But this is impossible:
Current estimates of the equilibrium climate sensitivity are of order
$0.8\,\mathrm{K}$ surface warming per $1\,\mathrm{W\,m^{-2}}$
radiative forcing at the top of the atmosphere, and the radiative
forcing at the surface can be of the same order as that at the top of
the atmosphere (though they are generally not equal). So a 3-K
global-mean surface temperature increase is inconsistent with a
$56$-$\mathrm{W\,m^{-2}}$ increase in net irradiance at the
surface. Likewise, the upward sensible heat flux cannot decrease
sufficiently to provide the additional energy flux at the surface
because it amounts to only about $20\,\mathrm{W\,m^{-2}}$ in the
global mean and $10\,\mathrm{W\,m^{-2}}$ in the mean over oceans,
where most evaporation occurs \citep{Kiehl97,Trenberth09a}.  The
implication of these order-of-magnitude arguments is that changes in
near-surface relative humidity and in evaporation---and thus, in a
statistically steady state, in global-mean precipitation---are
strongly energetically constrained. Order-of-magnitude estimates of
the climate sensitivity indicate global-mean evaporation can change by
$O(2\%\,\mathrm{K^{-1}})$, and the relation \eqref{e:rh_change} then
implies that the near-surface relative humidity can change by
$O(1\%\,\mathrm{K^{-1}})$ or less.

\begin{figure}[!tbh]
  \noindent\centerline{\includegraphics{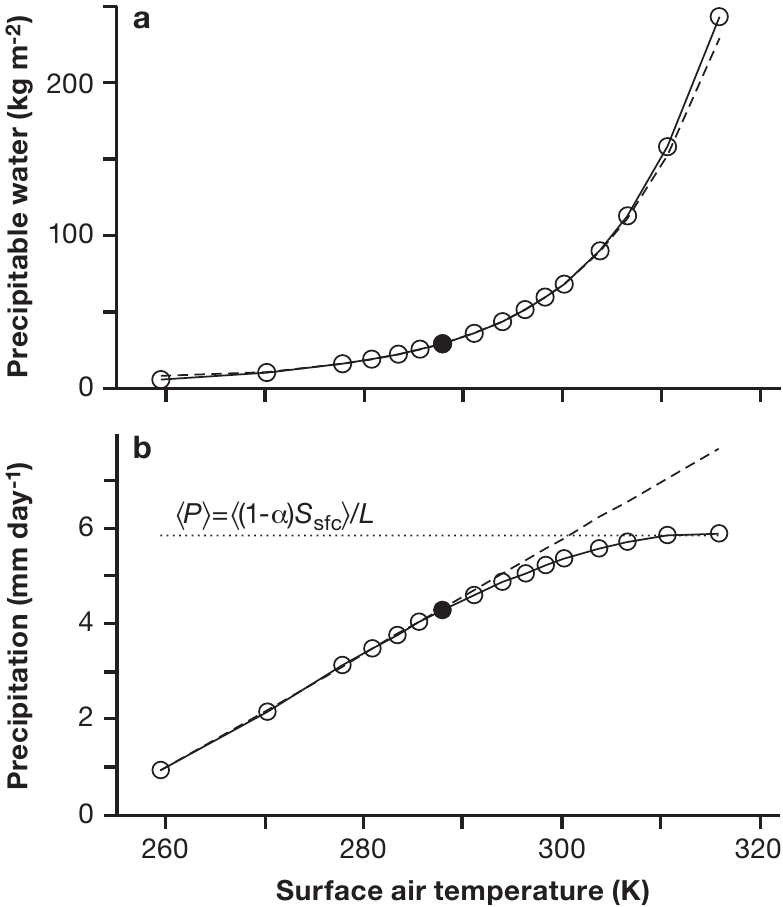}}
  \caption{Global-mean precipitable water and precipitation vs
    global-mean surface temperature in idealized GCM simulations. Each
    circle represents a statistically steady state of a GCM
    simulation. The filled circle marks a reference simulation with a
    climate resembling that of present-day Earth.  (a) Precipitable
    water. The dashed line is the global-mean column-integrated
    saturation specific humidity, calculated excluding levels in the
    upper atmosphere (pressures $\lesssim 0.05\,\mathrm{hPa}$) and
    rescaled by a constant effective relative humidity factor of
    $0.67$. In the idealized GCM, the specific latent heat of
    vaporization is taken to be constant, and the saturation specific
    humidity is calculated consistently with this approximation. (b)
    Precipitation. The dashed line shows the approximate upper bound
    \eqref{e:Pmax}. (Adapted from \citet{OGorman08b}.)}
  \label{f:wv_precip}
\end{figure}
The expectations based on the energetic arguments are borne out in the
idealized GCM simulations mentioned in the introduction and described
in \citet{OGorman08b}. Over a broad range of climates and in the
global mean, precipitable water increases exponentially with surface
temperature, roughly at the same rate as the column-integrated
saturation specific humidity, which is dominated by near-surface
contributions (Fig.~\ref{f:wv_precip}a). The effective relative
humidity varies little with climate, compared with the variations in
precipitable water by almost two orders of magnitude.  Nonetheless,
the effective relative humidity is not exactly constant but increases
by about 5 percentage points from the colder to the warmer simulations
(if the stratosphere is excluded from the calculation of the
column-integrated saturation specific humidity, otherwise the increase
is larger, and it generally is sensitive to precisely how it is
calculated).  The increase in the effective relative humidity is
qualitatively consistent with the relation \eqref{e:rh_change}, which
implies an increase in the near-surface relative humidity if the
fractional increase in saturation specific humidity exceeds that in
evaporation, as it does in all but the coldest simulations. However,
the energetic arguments only constrain the near-surface relative
humidity, not the relative humidity of the free troposphere. The
latter varies substantially among the simulations. For example, in the
extratropical free troposphere, the relative humidity decreases by
more than 15 percentage points from the coldest to the warmest
simulation \citep[][their Fig.~1]{OGorman08b}. The relative humidity
also changes more strongly in the free troposphere than near the
surface in simulations of climate change scenarios with comprehensive
GCMs \citep{Lorenz07b}. Contrary to what is sometimes surmised, there
is no universal principle that constrains free-tropospheric relative
humidity changes to be negligible or even to be of the same sign as
near-surface relative humidity changes. This implies in particular
that the energetic arguments alone do not constrain the strength of
the radiative water vapor feedback, which is sensitive to the
free-tropospheric specific humidity \citep[e.g.,][]{Held00a}.

In cold and moderately warm simulations, precipitation increases
roughly linearly with surface temperature in the global mean and
asymptotes to an approximately constant value in the warmest
simulations (Fig.~\ref{f:wv_precip}b). Precipitation generally
increases more slowly with surface temperature than does precipitable
water, except in the coldest simulations. For example, at the
reference simulation with global-mean surface temperature closest to
that of present-day Earth (288~K, filled circle in
Fig.~\ref{f:wv_precip}), precipitable water increases at
$6.2\%\,\mathrm{K^{-1}}$ in the global mean, whereas precipitation
increases at only $2.5\%\,\mathrm{K^{-1}}$. It is unclear why
precipitation increases roughly linearly with surface temperature over
a wide range of climates; energetic constraints appear to play a role
\citep{OGorman08b}. The constant value to which the precipitation
asymptotes is that at which the solar radiation absorbed at the
surface approximately balances the latent heat flux and thus
evaporation and precipitation in the global mean,
\begin{equation}\label{e:Pmax}
\langle P\rangle_{\max} \approx \langle(1-\alpha)  S_{\mathrm{sfc}}\rangle/L.
\end{equation}
Here, $\langle\cdot\rangle$ denotes a global mean; $\alpha$ is the
surface albedo and $S_{\mathrm{sfc}}$ the downwelling solar radiative
flux at the surface, which both are fixed in our idealized GCM
simulations (in reality they would vary with climate because, e.g.,
the cloud albedo and the absorption of solar radiation by water vapor
would vary). In fact, the global-mean precipitation exceeds the value
\eqref{e:Pmax} slightly in the warmest simulations because in warm
climates there is a net sensible heat flux from the atmosphere to the
surface \citep{Pierrehumbert02}. The sensible heat flux adds to the
absorbed solar irradiance in providing energy available to evaporate
water. (The net of the upwelling and downwelling longwave radiative
fluxes is small in the warmest simulations with atmospheres that are
optically thick for longwave radiation.)

The simulation results make explicit how the energy balance constrains
changes in precipitable water and precipitation. It should be borne in
mind that the energetic arguments constrain only the relative humidity
near the surface, not in the free atmosphere, and only the global-mean
precipitation and evaporation, not local precipitation, which is
influenced by transport of water vapor in the atmosphere. Local
precipitation may increase more rapidly with surface temperature than
global-mean precipitation, as may have happened in the past decades
over parts of the tropics (e.g., over oceans)
\citep{Gu07,Allan07}. However, reports that global precipitation and
evaporation increase much more rapidly with surface temperature than
stated here \citep[e.g.,][]{Wentz07} have to be regarded with caution;
they may be affected by measurement and analysis errors and
uncertainties resulting from estimating trends from noisy time series
\citep[see also][]{Adler08, Stephens08a}.

\section{Tropical circulations}\label{s:tropics}

\subsection{Gross Upward Mass Flux}

\begin{figure}[!tbh]
  \noindent\centerline{\includegraphics{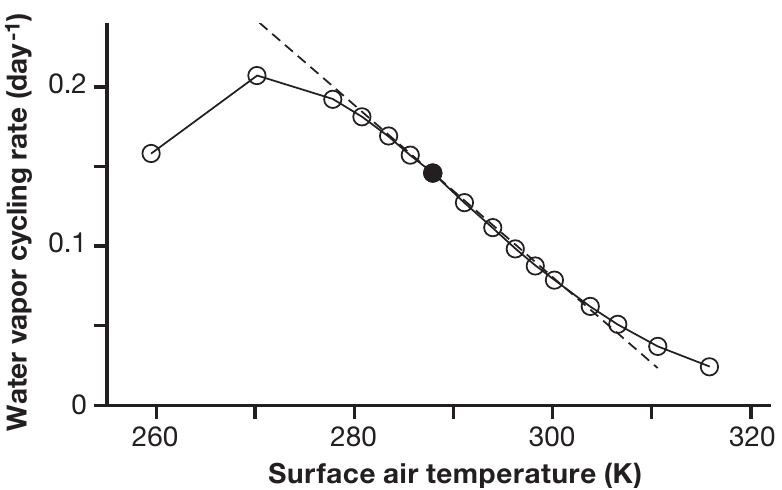}}
  \caption{Water vapor cycling rate vs global-mean surface temperature
    in idealized GCM simulations. The cycling rate is the ratio of
    global-mean precipitation (Fig.~\ref{f:wv_precip}b) and
    precipitable water (Fig.~\ref{f:wv_precip}a, up to a factor of
    water density). The dashed line marks a decrease of cycling rate
    of $3.7\% \,\mathrm{K^{-1}}$ relative to the reference simulation
    (filled circle).}
  \label{f:cycling_rate}
\end{figure}

That global-mean precipitable water and precipitation change with
climate at different rates has one immediate consequence: the water
vapor cycling rate---the ratio of global-mean precipitation and
precipitable water---changes. Global-mean precipitation increases more
slowly with surface temperature than does global-mean precipitable
water for all but the two coldest idealized GCM simulations. Hence,
the water vapor cycling rate decreases with surface temperature for
all but the two coldest simulations, from more than
$0.15\,\mathrm{day^{-1}}$ in the colder simulations to less than
$0.025\,\mathrm{day^{-1}}$ in the warmest simulations
(Fig.~\ref{f:cycling_rate}). At the reference simulation, the water
vapor cycling rate decreases with global-mean surface temperature at
$3.7\%\,\mathrm{K^{-1}}$, the difference between the rates of increase
in precipitation ($2.5\%\,\mathrm{K^{-1}}$) and precipitable water
($6.2\%\,\mathrm{K^{-1}}$). The water vapor cycling rate decreases at
similar rates in simulations of climate change scenarios with
comprehensive GCMs
\citep[e.g.,][]{Knutson95,Roads98,Bosilovich05,Held06,Stephens08a}. A
decreasing water vapor cycling rate may be interpreted as a weakening
of the atmospheric water cycle and may imply a weakening of the
atmospheric circulation, particularly in the tropics where most of the
water vapor is concentrated and precipitation is maximal
\citep[e.g.,][]{Betts89,Betts98,Held06,Vecchi06,Vecchi07b}.

A more precise relation between precipitation, specific humidity, and
the gross upward (convective) mass flux in the tropics follows from
considerations of the water vapor budget. In updrafts in the tropical
troposphere, above the lifted condensation level where the updraft air
is saturated with water vapor, the dominant balance in the water vapor
budget is between vertical advection of water vapor and
condensation. That is,
\begin{equation}\label{e:updraft_moisture}
  - \omega^\uparrow \partial_p q^* \approx c,
\end{equation}
where $q^*$ is the saturation specific humidity, $c$ is the condensation
rate, and
\begin{equation}
  \omega^\uparrow = 
  \begin{cases}
    \omega & \text{if } \omega < 0\\
    0 & \text{if } \omega \ge 0
    \end{cases}
\end{equation}
is the upward component of the vertical velocity $\omega = Dp/Dt$ in
pressure coordinates. Integrating in the vertical yields a relation
between the upward velocity, precipitation, and saturation specific
humidity,
\begin{equation}\label{e:trop_precip}
  - \{ \omega^\uparrow \partial_p q^* \} \approx P,
\end{equation}
where $\{ \cdot \} = g^{-1} \int dp \, (\cdot)$ denotes the
mass-weighted vertical integral over an atmospheric column
\citep[cf.][chapter~9.14]{Iribarne81}. We have assumed that the
vertically integrated condensation rate is approximately equal to the
precipitation rate, $\{ c \} \approx P$, which means that we have
neglected evaporation or sublimation of condensate. This is
justifiable if the relation \eqref{e:trop_precip} is understood as
applying to horizontal averages over convective systems, such that the
upward velocity $\omega^\uparrow$ is the net upward velocity within
convective systems---the net of convective updrafts and convective
downdrafts induced by evaporation or sublimation of condensate. When
understood in this way, the relation \eqref{e:trop_precip} holds
instantaneously, not only in long-term averages, and can be used, for
example, to relate precipitation extremes to updraft velocities and
thermodynamic conditions, even in the extratropics
\citep{OGorman09a,OGorman09b}.

From the relation \eqref{e:trop_precip}, one can obtain different
scaling estimates that give qualitatively different predictions of how
the tropical gross upward mass flux changes with climate. If the bulk
of the condensation occurs between a near-surface level with
saturation specific humidity $q^*_s$ and some tropospheric level with
saturation specific humidity $q^*$, the gross upward mass flux scales
as
\begin{subequations}
  \begin{equation}\label{e:up_scale_a}
    -\frac{\omega^\uparrow}{g} \sim \frac{P}{\Delta q^*},
  \end{equation}
  where $\Delta q^* = q^*_s - q^*$. This scaling estimate was suggested
  by \citet{Betts98}, based on the radiative-convective equilibrium
  model of \citet{Betts89}. If one follows these authors or
  \citet{Held06} further and assumes that the relevant tropospheric
  saturation specific humidity $q^*$ either is negligible or scales
  linearly with the near-surface saturation specific humidity $q^*_s$,
  the estimate \eqref{e:up_scale_a} simplifies to
  \begin{equation}\label{e:up_scale_b}
    -\frac{\omega^\uparrow}{g} \sim \frac{P}{q^*_s}.
  \end{equation}
\end{subequations}
To the extent that global-mean precipitation and precipitable water
scale with the tropical precipitation and near-surface saturation
specific humidity (which is not guaranteed), this scaling estimate
implies that the tropical gross upward mass flux scales with the water
vapor cycling rate, as suggested by \citet{Held06}.

\begin{figure}[!tbh]
  \noindent\centerline{\includegraphics{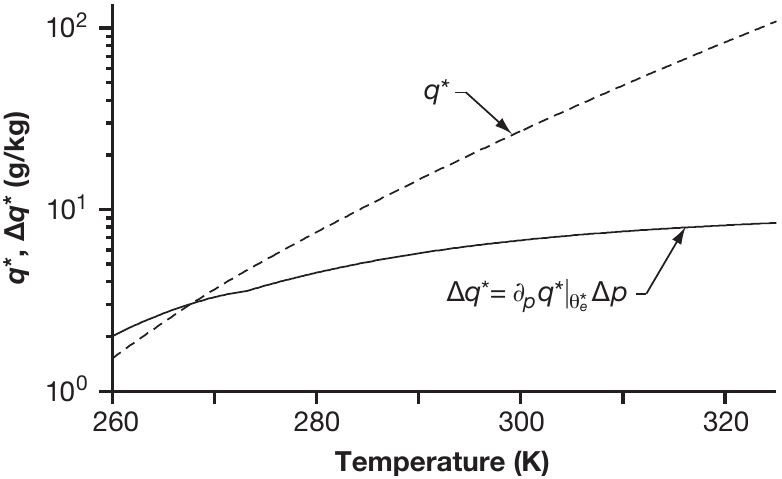}}
  \caption{Saturation specific humidity $q^*$ and saturation specific
    humidity contrast $\Delta q^* = \partial_p q^*|_{\theta_e^*} \,
    \Delta p$ as a function of temperature. Both are evaluated at
    $825\,\mathrm{hPa}$, and the pressure difference $\Delta p =
    250\,\mathrm{hPa}$ is taken to be fixed. The saturation specific
    humidity is calculated according to the modified Tetens formula
    given in \citet{Simmons99}, using the saturation vapor pressure
    over ice at very low temperatures, that over liquid water at
    temperatures above the freezing point, and a quadratic
    interpolation between the two at intermediate (mixed-phase)
    temperatures below the freezing point. (That is, freezing of water
    is taken into account in this figure, in contrast to the idealized
    GCM simulations, in which only the vapor-liquid phase transition
    is taken into account.) The fractional rate of increase of $q^*$
    varies between $9.5$ and $5.2\%\,\mathrm{K^{-1}}$ from low to high
    temperatures in the range shown, and that of $\Delta q^*$ varies
    between $6.6$ and $0.6\%\,\mathrm{K^{-1}}$. Note the logarithmic
    scale of the ordinate.}
  \label{f:qs_dpqs_change}
\end{figure}

The scaling estimates \eqref{e:up_scale_a} and \eqref{e:up_scale_b}
for the gross upward mass flux can differ substantially because the
saturation specific humidity contrast $\Delta q^*$ generally increases
less rapidly with temperature than the saturation specific humidity
$q^*$. For example, if the thermal stratification in convective
systems is moist adiabatic, the saturation specific humidity contrast
may scale as $\Delta q^* \sim \partial_p q^*|_{\theta_e^*} \Delta p$,
where the saturation specific humidity derivative is taken along a
moist adiabat with constant equivalent potential temperature
$\theta_e^*$, and the pressure difference $\Delta p$ is taken to be
fixed \citep{Betts87,OGorman09a,OGorman09b}. The saturation specific
humidity contrast $\Delta q^*$ then scales with the moist-adiabatic
static stability $S^* = -(T/\theta) \partial_p \theta|_{\theta_e^*}$
(potential temperature $\theta$) because on a moist adiabat, adiabatic
cooling balances diabatic heating through latent heat release, so that
the static stability and saturation specific humidity derivative are
related by $S^* \approx (L/c_p) \partial_p q^*|_{\theta_e^*}$
\citep[e.g.,][chapter~7.8]{Iribarne81}.
Now, the saturation specific humidity contrast $\Delta q^*$ generally
increases with temperature at a smaller fractional rate than the
saturation specific humidity $q^*$, with the difference between the
rates increasing with temperature (Fig.~\ref{f:qs_dpqs_change}). At a
temperature and pressure typical of the tropical lower troposphere in
the present climate ($290\,\mathrm{K}$ and $825\,\mathrm{hPa}$),
$\Delta q^*$ increases with temperature at $2.0\%\,\mathrm{K^{-1}}$,
while $q^*$ increases at $6.4\%\,\mathrm{K^{-1}}$. A fractional
increase in tropical precipitation of $2.5\%\,\mathrm{K^{-1}}$
(relative to a lower-tropospheric temperature) would imply a change in
the gross upward mass flux of $(2.5 - 2.0)\% \,\mathrm{K^{-1}}=
0.5\%\,\mathrm{K^{-1}}$ according to the estimate \eqref{e:up_scale_a}
but of $(2.5 - 6.4)\% \,\mathrm{K^{-1}}= -3.9\%\,\mathrm{K^{-1}}$
according to the estimate \eqref{e:up_scale_b}. Thus, the differences
between the two estimates can imply changes in the gross upward mass
flux of opposite sign: slight strengthening according to
\eqref{e:up_scale_a} and weakening according to
\eqref{e:up_scale_b}. Both estimates are based on rough scaling
assumptions, and neither may be very accurate (e.g., the relevant
pressure difference $\Delta p$ is not necessarily fixed but may vary
with climate). But they illustrate that the gross upward mass flux
does not necessarily scale with the water vapor cycling rate and may
depend, for example, on the vertical profile of the upward velocity
(averaged over convective systems, that is, including contributions
from convective downdrafts).

\begin{figure}[!tbh]
  \noindent\centerline{\includegraphics{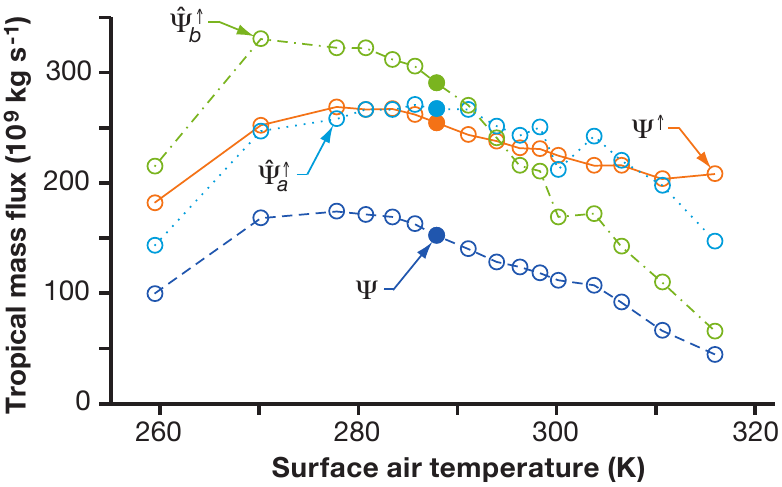}}
  \caption{Tropical vertical mass flux and scaling estimates vs
    global-mean surface temperature in idealized GCM
    simulations. Shown are the integrated gross upward mass flux
    $\Psi^\uparrow$, the mass transport streamfunction $\Psi$, and the
    scaling estimates $\hat\Psi^\uparrow_a$ and $\hat\Psi^\uparrow_b$
    corresponding to Eqs.~\eqref{e:up_scale_a} and
    \eqref{e:up_scale_b}, all evaluated at $4^\circ$ latitude and at a
    pressure of approximately $825\,\mathrm{hPa}$ and averaged over
    both, statistically identical, hemispheres. The scaling estimates
    $\hat\Psi^\uparrow_a$ and $\hat\Psi^\uparrow_b$ are multiplied by
    constants ($2.6$ and $1.6$, respectively) that are chosen such
    that the mean-square deviation between the scaling estimate and
    the integrated gross upward mass flux $\Psi^\uparrow$ is
    minimized.}
  \label{f:gcm_upward_mass}
\end{figure}

We test the scaling estimates for the tropical gross upward mass flux
using the upward mass flux on the idealized GCM's grid scale, sampled
four times daily, as a proxy for the unresolved subgrid-scale
convective mass flux.\footnote{The parameterized convection in the
  idealized GCM does not contain an explicit representation of subgrid
  convective mass fluxes. It acts by imposing temperature and specific
  humidity tendencies, as in the Betts-Miller convection scheme
  \citep{Betts86a,Betts86b,Betts93}. Vertical motion on the grid scale
  is induced by the thermodynamic effects of the parameterized
  convection, so it consists of convective and (particularly in the
  extratropics) large-scale components.} We have verified that the
grid-scale upward mass flux satisfies relation \eqref{e:trop_precip},
so that any errors in the scaling estimates are due to the assumptions
made in the estimates. Integrating the grid-scale upward mass flux
over an equatorial latitude band gives
\begin{equation}\label{e:psi_up}
  \Psi^\uparrow(\phi, p) = - \frac{2\pi a^2}{g} \int_{0^\circ}^{\phi}
  \Bar\omega^\uparrow(\phi', p) \cos\phi' \, d\phi', 
\end{equation}
where $a$ is Earth's radius, $\phi$ is latitude, and the overbar
denotes a zonal and temporal mean along isobars. With these
conventions, the integrated gross upward mass flux $\Psi^\uparrow$ is
directly comparable with the (net) mass transport streamfunction
$\Psi$, which, because the simulations are statistically symmetric
about the equator, is obtained by replacing the upward velocity
$\omega^\uparrow$ in \eqref{e:psi_up} with the net vertical velocity
$\omega$. Figure~\ref{f:gcm_upward_mass} shows $\Psi^\uparrow$ and
$\Psi$ evaluated at $4^\circ$ latitude and at a pressure\footnote{More
  precisely, Fig.~\ref{f:gcm_upward_mass} shows $\Psi^\uparrow$ and
  $\Psi$ defined analogously to \eqref{e:psi_up} but in $\sigma$
  coordinates and evaluated at $\sigma=0.825$, where $\sigma=p/p_s$
  (pressure $p$ over surface pressure $p_s$) is the GCM's vertical
  coordinate. In what follows, all quantities calculated from GCM
  simulations and reanalyses are evaluated in $\sigma$ coordinates,
  with the appropriate surface pressure-weighting of averages
  \citep[e.g.,][]{Walker06}; however, we give approximate pressure
  levels and expressions in pressure coordinates to simplify the
  presentation.} of approximately $825\,\mathrm{hPa}$ (i.e., it shows
mass fluxes across the 825-hPa level integrated between the equator
and $4^\circ$). The 825-hPa level is in all simulations within
$\lesssim 50\,\mathrm{hPa}$ of the level at which the gross upward
mass flux is maximal and at which the condensation in the column can
be expected to be maximal. (The level of maximum gross upward mass
flux likely depends on specifics of the convection and radiation
parameterization and so may be different in other GCMs.) The figure
also shows the estimates $\hat\Psi^\uparrow_a$ and
$\hat\Psi^\uparrow_b$ for the integrated gross upward mass flux that
are obtained by substituting the scaling estimates
\eqref{e:up_scale_a} and \eqref{e:up_scale_b} for the upward mass flux
$-\omega^\uparrow/g$ in \eqref{e:psi_up}. We evaluate the near-surface
saturation specific humidity $q_s^*$ at $950\,\mathrm{hPa}$ and the
tropospheric saturation specific humidity $q^*$ at
$700\,\mathrm{hPa}$---levels chosen to fit the estimates to the actual
gross upward mass flux as closely as possible. It is evident that the
estimate $\hat\Psi^\uparrow_b$ overestimates the changes in the gross
upward mass flux.  The water vapor cycling rate in
Fig.~\ref{f:cycling_rate} scales similarly to the estimate
$\hat\Psi^\uparrow_b$, so it likewise is not a good estimate of the
gross upward mass flux. The estimate $\hat\Psi^\uparrow_a$ gives a
better fit. At the reference simulation, the gross upward mass flux
decreases with global-mean surface temperature at about $1 \% \,
\mathrm{K^{-1}}$---more slowly by a factor $\about 3$ than the
estimate $\hat\Psi^\uparrow_b$ or the water vapor cycling rate, and
roughly consistent with the moist-adiabatic static stability arguments
and the estimate $\hat\Psi^\uparrow_a$ (Fig.~\ref{f:qs_dpqs_change}).

The simulations demonstrate that at least in this idealized GCM, to
understand changes in the gross upward mass flux, it is important to
consider not just changes in the near-surface saturation specific
humidity but changes in the saturation specific humidity
stratification, or in the static stability---as did, for example,
\citet{Knutson95}. The corresponding scaling estimates are clearly
distinguishable. They not only imply quantitatively different rates at
which the gross upward mass flux changes with climate; they can also
imply qualitatively different results in that their maxima occur in
different climates (Fig.~\ref{f:gcm_upward_mass}). Because the gross
upward mass flux in the tropics represents the bulk of the global
gross upward mass flux, similar conclusions to those drawn here for
the tropics also apply to the global mean.

The gross upward mass flux in Earth's tropical atmosphere appears to
have decreased as the climate warmed in recent decades
\citep{Tanaka04,Vecchi06,Zhang06}. These observations are consistent
with the idealized GCM simulations, in which the gross upward mass
flux in the tropics exhibits a maximum at a climate somewhat colder
than that of the present day. By how much the gross upward mass flux
in Earth's tropical atmosphere flux has decreased, however, is
difficult to ascertain because of data uncertainties. In simulations
of climate change scenarios, the gross upward mass flux also decreases
as the surface temperature increases, both globally and in the
tropics, with most of the decrease in the tropics occurring in zonally
asymmetric circulation components (e.g., in the Walker circulation),
not in the zonal-mean Hadley circulation \citep{Held06}. The gross
upward mass flux decreases more slowly than the water vapor cycling
rate in almost all models used in the IPCC Fourth Assessment Report
\citep{Vecchi07b}. \citet{Vecchi07b} interpreted this result as being
roughly consistent with the scaling of the gross upward mass flux with
the water vapor cycling rate and speculated that non-precipitating
upward mass fluxes are responsible for the systematic deviations from
this scaling. However, their results appear more consistent with our
idealized GCM simulations and with the assumption that the saturation
specific humidity stratification, rather than the water vapor cycling
rate, is important for the scaling of the gross upward mass
flux.\footnote{\citet{Held06} and \citet{Vecchi07b} consider the
  mid-tropospheric mass flux at $500\,\mathrm{hPa}$, rather than a
  lower-tropospheric mass flux. Generally they find that the
  mid-tropospheric gross upward mass flux on the grid scale of the
  models decreases more slowly than the water vapor cycling rate as
  the climate warms. In one model, however, the mid-tropospheric
  convective mass flux scales with the water vapor cycling rate at
  least over the earlier part of a 21st-century climate change
  simulation (it varies more slowly in later parts of the
  simulation). But this latter result may be not be general. In our
  idealized GCM simulations, the mid-tropospheric gross upward mass
  flux also scales with the water vapor cycling rate near the
  reference simulation and in warmer simulations, but not in colder
  simulations (excluding the coldest simulation, in which the
  tropopause is near or below the $500$-hPa level). The
  mid-tropospheric gross upward mass flux is up to a factor $\about 2$
  smaller than the lower-tropospheric gross upward mass flux we
  consider, so the latter may be more clearly related to
  column-averaged condensation and precipitation, at least in our
  GCM.}

Thus, in climates similar to the present or warmer, the gross upward
mass flux in the tropics likely decreases as the climate
warms. Convective activity, by this bulk measure, likely decreases as
the climate warms---which may seem counterintuitive because it
generally increases with surface temperature (or near-surface specific
humidity) when spatial or temporal fluctuations within the present
climate are considered. The reason for the different responses is that
water vapor dynamics plays different roles in climate changes and in
fluctuations within a given climate. As the climate warms, when
surface temperatures increase on large scales, large-scale
precipitation changes are energetically constrained, latent heat
release in moist convection increases the large-scale tropical static
stability (the moist-adiabatic lapse rate decreases), and both effects
together can lead to a weakening of the gross upward mass flux
\citep{Betts98}. In fluctuations within a given climate, the static
stability is controlled by processes on large scales, and latent heat
release can locally induce potentially strong upward mass fluxes. This
illustrates how misleading it can be to use fluctuations within the
present climate (such as El Ni{\~n}o and the Southern Oscillation) for
inferences about climate changes. For example, while observations
suggest that there may be a threshold sea surface temperature that
must be exceeded for strong convection to occur over Earth's tropical
oceans \citep[e.g.,][]{Graham87,Folkins03}, there is no justification
for using the same threshold temperature for inferences about
convection in changed climates: to the extent that such a threshold
temperature exists, it may change as the climate changes, and with it
the large-scale tropical static stability
\citep[e.g.,][]{Knutson95,Neelin09a}.

Our focus has been on integrated measures of the gross upward mass
flux, which are constrained by large-scale energetic and hydrologic
balances. Regionally, the response to climate changes is less
constrained and can be more complex. For example, margins of
convective regions are particularly susceptible to relatively large
changes in upward mass fluxes and precipitation
\citep[e.g.,][]{Neelin03,Chou04,Neelin06,Neelin07a,Chou09}.

\subsection{Strength of Hadley Circulation}\label{s:hadley_strength}

While arguments based on energetic and hydrologic balances alone
constrain how the tropical gross upward mass flux changes with
climate, they are generally insufficient to constrain how the net
vertical mass flux and thus the strength of the Hadley circulation
change. Even near the equator, within the ascending branch of the
Hadley circulation, the net vertical mass flux amounts to only a
fraction of the gross upward mass flux. For example, in the idealized
GCM simulations, the gross upward mass flux $\Psi^\uparrow$ in the
lower troposphere, integrated over an equatorial latitude band within
the ascending branch of the Hadley circulation, is a factor $2$--$5$
larger than the corresponding net vertical mass flux $\Psi$
(Fig.~\ref{f:gcm_upward_mass}). This means that even in this
equatorial latitude band, $1/2$ to $4/5$ of the upward mass fluxes is
offset by downward mass fluxes between the (parameterized) convective
systems in which the upward mass fluxes occur. In the idealized GCM
simulations, the net vertical mass flux $\Psi$ scales similarly to the
gross upward mass flux $\Psi^\uparrow$, except in the warmest
simulations (Fig.~\ref{f:gcm_upward_mass}), but this is not generally
so; we have obtained simulations with an idealized GCM containing a
representation of ocean heat transport in which the two mass fluxes
scale differently over a broad range of climates. 

The reason why the strength of the Hadley circulation responds
differently to climate changes than the gross upward mass flux is that
the Hadley circulation is not only constrained by energetic and
hydrologic balances but also by the angular momentum balance, which it
must obey irrespective of water vapor dynamics. In the upper
troposphere above the center of the Hadley cells---where frictional
processes and the vertical advection of momentum by the mean
meridional circulation are negligible---the mean balance of angular
momentum about Earth's spin axis in a statistically steady state is
approximately
\begin{equation}\label{e:zon_mom}
  (f + \Bar\zeta) \Bar v = f(1-\Ro) \Bar v \approx \mathcal{S}.
\end{equation}
Here, $\Ro = -\Bar\zeta/f$ is a spatially varying local Rossby number
with Coriolis parameter $f$ and relative vorticity $\zeta$, $v$ is the
meridional velocity, and $\mathcal{S}$ is the eddy (angular) momentum
flux divergence \citep{Schneider06b,Walker06}. The Hadley circulation
conserves angular momentum in its upper branch in the limit $\Ro \to
1$ and $\mathcal{S} \to 0$, in which the angular momentum or zonal
momentum balance \eqref{e:zon_mom} degenerates and provides no
constraint on the mean meridional mass flux ($\propto \Bar v$). Only
in this limit does the Hadley circulation strength respond directly to
changes in thermal driving \citep[cf.][]{Held80}. In the limit $\Ro
\to 0$, the Hadley circulation strength ($\propto \mathcal{S}/f$)
responds to climate changes only via changes in the eddy momentum flux
divergence $\mathcal{S}$, and possibly via changes in the width of the
Hadley cells that can affect the relevant value of the Coriolis
parameter $f$ near their center. In this limit, changes in thermal
driving affect the Hadley circulation strength only insofar as they
affect the eddy momentum flux divergence $\mathcal{S}$ or the relevant
value of the Coriolis parameter $f$. The local Rossby number $\Ro$
above the center of a Hadley cell is a nondimensional measure of how
close the upper branch is to the angular momentum-conserving limit. In
the limit $\Ro \to 1$, nonlinear momentum advection by the mean
meridional circulation, $f \Ro \, \Bar v = \Bar v (a
\cos\phi)^{-1} \partial_{\phi} (\Bar u \cos\phi)$, dominates over eddy
momentum flux divergence.  In the limit $\Ro \to 0$, eddy momentum
flux divergence dominates over nonlinear momentum advection by the
mean meridional circulation.

For intermediate local Rossby numbers $0 < \Ro < 1$, the Hadley
circulation strength can respond to climate changes both via changes
in the eddy momentum flux divergence and via changes in the local
Rossby number. The zonal momentum balance \eqref{e:zon_mom} implies
that a small fractional change $\delta \Bar v/\Bar v$ in the strength
of the upper-tropospheric mean meridional mass flux must be met by
changes in the eddy momentum flux divergence, $\delta \mathcal{S}$, in
the local Rossby number, $\delta \Ro$, and in the relevant value of
the Coriolis parameter, $\delta f$, satisfying
\begin{equation}
  \frac{\delta \Bar v}{\Bar v} 
  \approx \frac{\delta \mathcal{S}}{\mathcal{S}} 
  + \frac{\delta\Ro}{1-\Ro} - \frac{\delta f}{f}. 
\end{equation}
For example, if $\Ro = 0.2$, and if we neglect changes in the relevant
value of the Coriolis parameter near the center of the Hadley cells, a
$10\%$ increase in the strength of the mean meridional mass flux
requires a $10\%$ increase in $\mathcal{S}$, or an increase in $\Ro$
of $\delta \Ro = 0.08$ or $40\%$, or a combination of these two kinds
of changes. A $40\%$ increase in $\Ro$ implies the same increase in
the relative vorticity (meridional shear of the zonal wind) and hence
a similarly strong increase in upper-tropospheric zonal winds. Such a
strong increase in zonal winds would almost certainly affect the eddy
momentum flux divergence $\mathcal{S}$ substantially. For example,
according to the scaling laws in \citet{Schneider08b}, the eddy
momentum flux divergence scales at least with the square root of
meridional surface temperature gradients and thus upper-tropospheric
zonal winds (by thermal wind balance). So for small $\Ro$ in general,
changes in $\mathcal{S}$ are strongly implicated in any changes in
Hadley circulation strength. Conversely, if $\Ro=0.8$ under the same
assumptions, a $10\%$ increase in the strength of the mean meridional
mass flux requires an increase in $\Ro$ of only $\delta \Ro = 0.02$ or
$2.5\%$---implying much subtler changes in upper-tropospheric zonal
winds with a weaker effect on eddy momentum fluxes. So for large $\Ro$
in general, changes in $\mathcal{S}$ play a reduced role in changes in
Hadley circulation strength, which therefore can respond more directly
to climate changes via changes in energetic and hydrologic balances.

\begin{figure*}[!tb]
  \noindent\centerline{\includegraphics{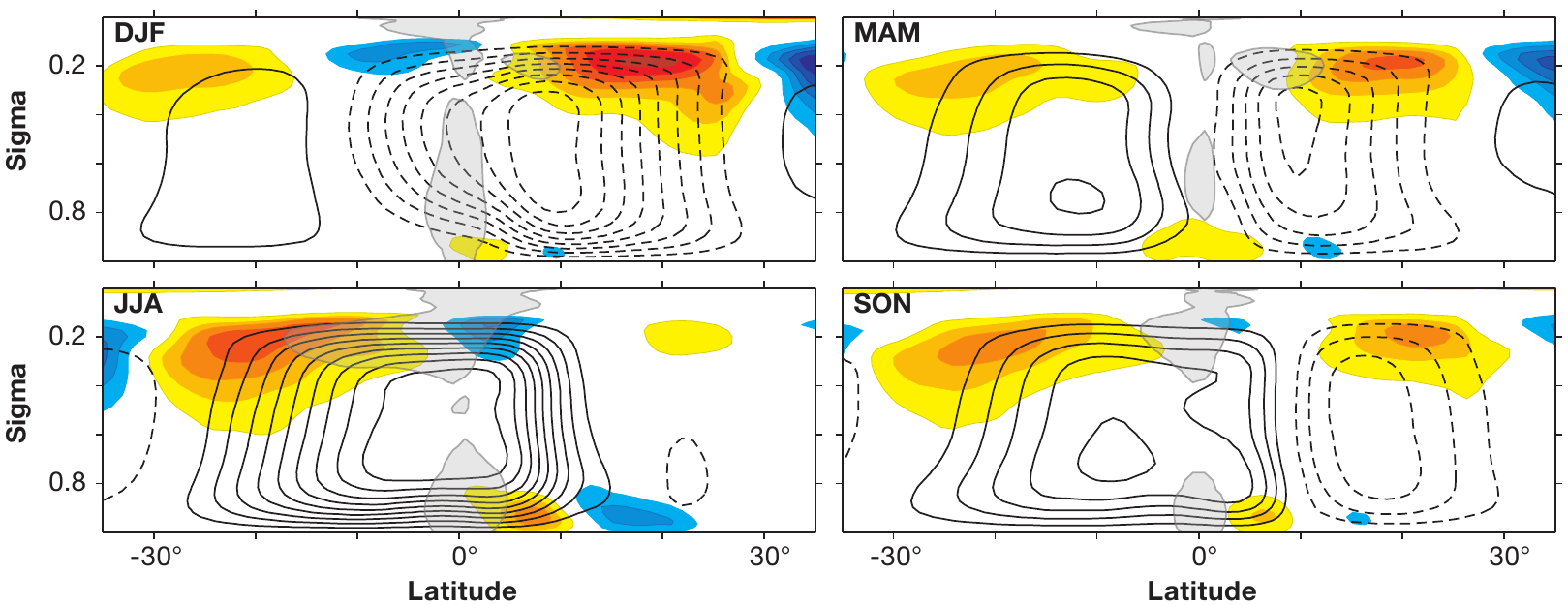}}
  \caption{Earth's Hadley circulation over the course of the seasonal
    cycle. Black contours show the mass flux streamfunction, with
    dashed (negative) contours indicating clockwise motion and solid
    (positive) contours indicating counterclockwise motion (contour
    interval $25\times10^9\,\mathrm{kg}\,\mathrm{s}^{-1}$). Colors
    indicate horizontal eddy momentum flux divergence
    $\mathop{\mathrm{div}}(\overline{u' v'} \cos\phi)$, with the
    overbar denoting the seasonal and zonal mean and primes denoting
    deviations therefrom (contour interval $8 \times
    10^{-6}\,\mathrm{m\,s^{-2}}$, with red tones for positive and blue
    tones for negative values). Gray shading indicates regions in
    which $| \Ro | > 0.5$. The vertical coordinate $\sigma = p/p_s$ is
    pressure $p$ normalized by surface pressure $p_s$. Computed from
    reanalysis data for the years 1980--2001 provided by the European
    Centre for Medium-Range Weather Forecasts
    \protect\citep{Kallberg04,Uppala05}.}
  \label{f:hadley_seasonal}
\end{figure*}

Earth's Hadley cells most of the year exhibit relatively small local
Rossby numbers in their upper branches, but local Rossby numbers and
the degree to which the Hadley cells are influenced by eddy momentum
fluxes vary over the course of the seasonal
cycle. Figure~\ref{f:hadley_seasonal} shows Earth's Hadley circulation
and the horizontal eddy momentum flux divergence for
December--January--February (DJF), March--April--May (MAM),
June--July--August (JJA), and September--October--November (SON). Also
shown are the regions in which $|\Ro| > 0.5$ in the zonal and seasonal
mean, that is, regions in which nonlinear momentum advection by the
mean meridional circulation is a dominant term in the zonal momentum
balance. It is evident that the strength of the DJF, MAM, and SON
Hadley cells in both hemispheres is primarily controlled by the eddy
momentum flux divergence ($|\Ro| < 0.5$ throughout much of their upper
branches above their centers, with $|\Ro| \ll 1$ in the summer and
equinox cells); only the strength of the cross-equatorial JJA Hadley
cell is not primarily controlled by the eddy momentum flux divergence
($\Ro| \gtrsim 0.5$ in much of its upper branch)
[\citealt{Walker05,Walker06,Schneider08a}; see \citealt{Dima05} for a
more detailed analysis of the tropical zonal momentum
balance]. Nonlinear momentum advection by the mean meridional
circulation is a dominant term in the zonal momentum balance in the
upper branch of the cross-equatorial JJA Hadley cell, which primarily
consists of the Asian summer monsoon circulation \citep{Dima03}.  In
the annual mean, Earth's Hadley cells have $|\Ro| < 0.5$ throughout
their upper branches, so their strength as well as that of the DJF,
MAM, and SON Hadley cells responds to climate changes primarily via
changes in the eddy momentum flux divergence. Consistent with these
observations, interannual variations in the strength of the DJF Hadley
cells are correlated with interannual variations in the eddy momentum
flux divergence \citep{Caballero07}, and differences in strength of
the DJF Hadley cells among climate models are correlated with
differences in the momentum flux divergence owing to stationary eddies
\citep{Caballero08}. However, the response of monsoonal circulations
to climate changes may be more directly controlled by energetic and
hydrologic balances and may differ from the response of the Hadley
cells during the rest of the year \citep{Bordoni08}. And while eddy
momentum fluxes constrain the strength of the Hadley cells, that is,
the streamfunction extremum at the center of the cells, they do not
necessarily constrain where the ascent occurs and thus the position of
the Intertropical Convergence Zone, as local Rossby numbers in the
ascending branches can be large (Fig.~\ref{f:hadley_seasonal}).

\begin{figure}[!tbh]
  \noindent\centerline{\includegraphics{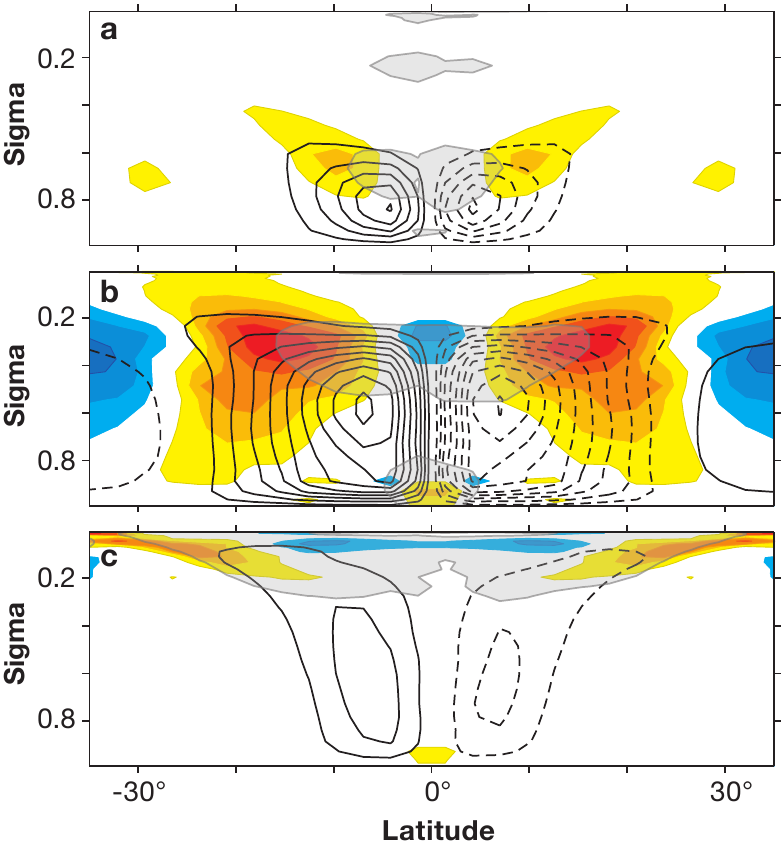}}
  \caption{Hadley circulations in three idealized GCM simulations. (a)
    Coldest simulation (global-mean surface temperature $\langle T_s
    \rangle = 259\,\mathrm{K}$). (b) Reference simulation ($\langle
    T_s \rangle = 288\,\mathrm{K}$). (c) Warmest simulation ($\langle
    T_s \rangle = 316\,\mathrm{K}$). Plotting conventions are as in
    Fig.~\ref{f:hadley_seasonal}, but with contour interval
    $20\times10^9\,\mathrm{kg}\,\mathrm{s}^{-1}$ for streamfunction
    (black) and $4 \times 10^{-6}\,\mathrm{m\,s^{-2}}$ for horizontal
    eddy momentum flux divergence (colors). Gray shading again
    indicates regions in which $| \Ro | > 0.5$.}
  \label{f:gcm_hadley}
\end{figure}

In the idealized GCM simulations, local Rossby numbers and the degree
to which eddy momentum fluxes influence the strength of the Hadley
circulation vary with climate. The Hadley circulation is generally
more strongly influenced by eddy momentum fluxes in colder climates
than in warmer climates: the local Rossby number in the upper branches
increases from $\lesssim 0.5$ in the coldest simulation to $\lesssim
0.8$ in the warmest simulation (Fig.~\ref{f:gcm_hadley}). So
understanding how the eddy momentum flux divergence in low latitudes
changes with climate is one important part of what needs to be
understood to explain how the strength of the Hadley circulation
changes in the simulations, but the nonlinear momentum advection by
the mean meridional circulation must also be taken into
account. However, the Hadley circulation in the simulations is
generally less strongly influenced by eddy momentum fluxes than
Earth's Hadley cells during equinox or in the annual mean (cf.\
Fig.~\ref{f:hadley_seasonal}). This is a consequence of neglecting
ocean heat transport, which dominates the meridional heat transport in
Earth's low latitudes \citep{Trenberth01}. Neglecting it leads to
stronger Hadley cells---e.g., by about 60\% in our reference
simulation compared with Earth's equinox or annual mean (compare
Figs.~\ref{f:hadley_seasonal} and \ref{f:gcm_hadley}), or by up to
$O(1)$ factors in simulations with comprehensive GCMs
\citep{Herweijer05,Lee08}. As a result, the nonlinear momentum
advection by the mean meridional circulation is stronger and the
Hadley cells are closer to the angular momentum--conserving limit than
Earth's Hadley cells. Neglecting the coupling of ocean heat transport
to the strength of the Hadley circulation \citep{Klinger00,Held01}
thus may lead to different responses of the Hadley circulation to the
seasonal cycle or to climate changes, as seen, for example, in the
simulations in \citet{Clement06} or \citet{Bliesner05}. Therefore, any
theory of how the Hadley circulation responds to climate changes must
build upon not only a theory of how eddy momentum fluxes change with
climate but also a theory of how ocean heat transport is coupled to
and modifies the Hadley circulation.

Like the tropical gross upward mass flux, the strength of the Hadley
circulation in the idealized GCM simulations changes non-monotonically
with global-mean surface temperature. The mass flux in the Hadley
cells is $104 \times 10^9\,\mathrm{kg\, s^{-1}}$ in the coldest
simulation, $184 \times 10^9\,\mathrm{kg\, s^{-1}}$ in the reference
simulation, and $51 \times 10^9\,\mathrm{kg\, s^{-1}}$ in the warmest
simulation (Fig.~\ref{f:gcm_hadley}). It is maximal in climates
slightly colder than that of present-day Earth (see
Fig.~\ref{f:gcm_upward_mass}, which shows the vertical mass flux
between the equator and $4^\circ$ at $825\,\mathrm{hPa}$, but this
closely approximates the strength of the Hadley cells, or the
streamfunction extremum). We have obtained qualitatively similar
behavior of the Hadley circulation strength in idealized GCM
simulations that do take coupling to ocean heat transport into account
\citep{Levine10}. This shows that the strength of the Hadley
circulation need not always decrease as the climate warms, although it
is plausible that it does so as the climate warms relative to that of
present-day Earth. Nonetheless, the Hadley circulation may also have
been weaker in much colder climates, such as that of the LGM or of a
completely ice-covered ``snowball'' Earth, which may have occurred
$\about 750$ million years ago \citep{Hoffman98}. However, the
presence of sea and land ice and ice-albedo feedbacks, which we
ignored, may modify the behavior seen in the idealized GCM
simulations.

Part of the reason for the non-monotonic change in Hadley circulation
strength with global-mean surface temperature is that eddy momentum
fluxes influence the Hadley circulation strength and themselves change
non-monotonically (see the color contours in
Fig.~\ref{f:gcm_hadley}). The eddy momentum flux divergence within the
Hadley circulation scales similarly to the extratropical eddy kinetic
energy \citep{Schneider08b}, which changes non-monotonically with
global-mean surface temperature (see Fig.~\ref{f:eke} below). The
reasons for the non-monotonic change in eddy kinetic energy will be
discussed further in Section~\ref{s:eke}. However, the changes in eddy
momentum flux divergence and eddy kinetic energy do not completely
account for the changes in Hadley circulation strength because local
Rossby numbers and the degree to which eddy momentum fluxes influence
the Hadley circulation vary with climate.

Because the strength of the Hadley circulation is partially controlled
by eddy momentum fluxes and extratropical eddy kinetic energies, it
bears no obvious relation to the tropical gross upward mass flux,
which is more directly controlled by energetic and hydrologic
balances. In general, reasoning about the strength of the Hadley
circulation that focuses on energetic and hydrologic balances alone
and does not take eddy momentum fluxes into account is likely going to
be inadequate, given the relatively small local Rossby numbers in
Earth's Hadley cells most of the year.

We currently do not have a theory of how the Hadley circulation
strength changes with climate. We have theories for the limit $\Ro \to
1$ \citep{SchneiderEK77b,Held80}, in which eddy momentum fluxes play
no role. We have theories for the limit $\Ro \to 0$
\citep[e.g.,][]{Dickinson71,SchneiderEK77a,Fang94}, in which nonlinear
momentum advection by the mean meridional circulation plays no role
and one needs primarily a theory of how eddy momentum fluxes change
with climate (such as presented in \citet{Schneider08b} for dry
atmospheres). What we need is a theory that can account for
interacting changes in the mean meridional circulation and in eddy
momentum fluxes, including changes in the relative importance of
nonlinear momentum advection by the mean meridional
circulation. Shallow-water models of the Hadley circulation in which
eddy effects are parameterized may be a starting point for the
development of such a theory \citep{Sobel09}.

Our discussion has focused on the eddy momentum flux as the primary
eddy influence on the strength of the Hadley circulation. Eddies can
also influence the strength of the Hadley circulation through their
energy transport \citep[e.g.,][]{Kim01b,Becker01}. For example, the
Hadley circulation is constrained by the requirement that diabatic
heating in the tropics balance cooling in the subtropics by both
radiative processes and eddy energy export to the
extratropics. However, unlike the momentum transport, the energy
transport by eddies throughout the bulk of Earth's Hadley cells is
smaller (albeit not by much) than the transport by the mean meridional
circulation, except in the descending branches
\citep{Trenberth03a}.\footnote{\citet{Trenberth03} argue that the
  ``seamlessness'' of the energy transport between the tropics and the
  extratropics provides a constraint on the energy transport and
  strength of Hadley cells. However, this ``seamlessness'' of the
  energy transport is not a fundamental property of atmospheric
  circulations. We have obtained simulations with idealized GCMs in
  which the tropical energy transport, dominated by the mean
  meridional circulation, differs substantially from the extratropical
  energy transport, dominated by eddies. (For example, this is the
  case in the simulations in \citet{Walker06} in which the planetary
  rotation rate is varied.)}  It can be incorporated relatively easily
into Hadley circulation theories as an additional thermal driving,
provided relations between the eddy energy transport and mean fields
such as temperature gradients can be established
\citep[e.g.,][]{Held80,SchneiderEK84,Schneider08b}.

\subsection{Height of Hadley Circulation}\label{s:tropopause}

Another change in the Hadley circulation evident in the idealized GCM
simulations is that its height, and with it the height of the tropical
tropopause, generally increases as the climate warms
(Fig.~\ref{f:gcm_hadley}). This can be understood from radiative
considerations (e.g., \citet{Held82,Thuburn00,Caballero08b}; see
\citet{Schneider07b} for a review).

A simple quantitative relation indicating how the tropical tropopause
height changes with climate can be obtained if (i) the tropospheric
temperature lapse rate is taken to be constant, (ii) the atmosphere is
idealized as semigray (transparent to solar radiation and gray for
longwave radiation), and (iii) the stratosphere is taken to be
optically thin and in radiative equilibrium (i.e., the effect of the
stratospheric circulation on the tropopause height is neglected). The
tropopause height $H_t$ is then related to the surface temperature
$T_s$, tropospheric lapse rate $\gamma$, and emission height $H_e$ (at
which the atmospheric temperature is equal to the emission
temperature) through
\begin{equation}\label{e:tropopause_height}
  H_t \approx (1-c)\frac{T_s}{\gamma} + c H_e,
\end{equation}
where $c=2^{-1/4}\approx 0.84$ \citep{Schneider07b}. As the
concentration of greenhouse gases (or the optical thickness of the
longwave absorber) increases, the emission height $H_e$ and the
tropical surface temperature $T_s$ generally increase. The tropical
lapse rate $\gamma$ generally decreases because it is close to the
moist-adiabatic lapse rate, which decreases with increasing
temperature. All three factors---increase in $T_s$, decrease in
$\gamma$, and increase in $H_e$---contribute to the increase in
tropopause height seen in the idealized GCM simulations.

The relation \eqref{e:tropopause_height} implies for a typical
tropical lapse rate of $6.5\,\mathrm{K\,km^{-1}}$ that an increase in
tropical surface temperature of $1\,\mathrm{K}$ leads to an increase
in tropopause height of $25\,\mathrm{m}$ if the emission height stays
fixed; any increase in the concentration of greenhouse gases implies
an increase in emission height, leading to an additional increase in
tropopause height (see \citet{Thuburn00} and \citet{Schneider07b} for
more precise calculations for more realistic atmospheres). Roughly
consistent with these arguments, the tropical tropopause height in
recent decades has increased by tens of meters \citep{Seidel01}, and
in simulations of climate change scenarios, it also increases with
tropical surface temperature at a rate of $\about
10$--$100\,\mathrm{m\,\mathrm{K^{-1}}}$
\citep{Santer03a,Santer03b,Bliesner05}.

\subsection{Width of Hadley Circulation}

The Hadley circulation appears to have widened in recent decades
\citep{Hu07b,Seidel07,Seidel08,Johanson09}, and it also widens, in the
annual mean, as surface temperatures increase in many simulations of
climate change scenarios \citep{Lu07}. How the width of the Hadley
circulation is controlled, however, is unclear.

Following the recognition that eddy fluxes are essential for the
general circulation \citep[e.g.,][]{Defant21,Jeffreys26} and can be
generated by baroclinic instability \citep{Charney47,Eady49}, it was
generally thought that the meridional extent of the Hadley circulation
is limited by baroclinic eddy fluxes. But the work of
\citet{SchneiderEK77b} and \citet{Held80} (and moist generalizations
such as that of \citet{Emanuel95c}) made clear that a Hadley
circulation even without eddy fluxes, with upper branches approaching
the angular momentum--conserving limit, does not necessarily extend to
the poles but can terminate at lower latitudes. The Hadley circulation
occupies the latitude band over which its energy transport needs to
extend to reduce meridional radiative-equilibrium temperature
gradients to values that are consistent with thermal wind balance and
with a zonal wind that does not violate the constraint of Hide's
theorem that there be no angular momentum maximum in the interior
atmosphere (\citet{Hide69}; \citet{Held80}; see \citet{Schneider06b}
for a review). \citet{Held80} calculated the strength and width of a
Hadley circulation under the assumptions that (i) the poleward flow in
the upper branches is approximately angular momentum--conserving and
(ii) the circulation is energetically closed, so that diabatic heating
in ascent regions is balanced by radiative cooling in descent
regions. In the small-angle approximation for latitudes and for
radiative-equilibrium temperatures that decrease quadratically with
latitude away from the equator (a good approximation for Earth in the
annual mean), the Hadley circulation according to the Held-Hou theory
extends to the latitude
\begin{equation}\label{e:phi_M}
  \phi_{\mathrm{HH}} \approx \left(\frac{5}{3} \, \frac{g H_t}{\Omega^2 a^2} \,
    \frac{\Delta_h}{T_0}\right)^{1/2},
\end{equation}
where $\Omega$ is the planetary angular velocity, $\Delta_h$ is the
(vertically averaged) pole-equator temperature contrast in radiative
equilibrium, and $T_0$ is a reference temperature. Substituting values
representative of Earth ($\Delta_h/T_0 \approx
80\,\mathrm{K}/295\,\mathrm{K}$ and $H_t \approx 15\,\mathrm{km}$)
gives the Hadley circulation terminus $\phi_{\mathrm{HH}} \approx
32^\circ$ (more precisely, $\phi_{\mathrm{HH}} = 29^\circ$ if the
small-angle approximation is not made). Because this latitude is
approximately equal to the actual terminus of the Hadley circulation
in Earth's atmosphere (cf.\ Fig.~\ref{f:hadley_seasonal}), the
Held-Hou result \eqref{e:phi_M} was subsequently often taken as
relevant for Earth's atmosphere. If applicable to Earth's atmosphere,
it would imply, for example, that the Hadley circulation widens as the
tropopause height or the pole-equator temperature contrast increase.

However, it is questionable how relevant \eqref{e:phi_M} is for the
response of Earth's Hadley circulation to climate changes. Because
Earth's Hadley circulation generally neither approaches the angular
momentum--conserving limit nor is it energetically closed
(section~\ref{s:hadley_strength}), it may respond differently to
climate changes. Indeed, even in simulations with an idealized dry
GCM, the width of the Hadley circulation does not behave as indicated
by \eqref{e:phi_M} in parameter regimes in which the Rossby number in
the circulation's upper branches is similar to that in Earth's
atmosphere \citep{Walker06}. For example, the Hadley circulation
widens much more slowly with increasing radiative-equilibrium
pole-equator temperature contrast than indicated by \eqref{e:phi_M};
it also widens with increasing low-latitude static stability, whereas
\eqref{e:phi_M} would imply it is independent of static stability.

The dependence of the width of the Hadley circulation on the
low-latitude static stability suggests a link to baroclinic eddy
fluxes. In dry atmospheres, an increased static stability means that
the latitude at which baroclinic eddy fluxes first become deep enough
to reach the upper troposphere moves poleward
\citep{Held78b,Schneider06a}. Therefore, it is plausible to attribute
the widening of the Hadley circulation with increasing low-latitude
static stability to a poleward displacement of deep baroclinic eddy
fluxes \citep{Walker06,Korty08}. Making this notion more precise and
harking back to earlier ideas about what terminates the Hadley
circulation, one may suppose that the Hadley circulation extends up to
the lowest latitude $\phi_e$ at which meridional eddy entropy fluxes
are deep enough to reach the upper troposphere \citep{Korty08}. At
this latitude, wave activity generated near the surface first reaches
the upper troposphere, as the meridional eddy entropy flux is
proportional to the vertical wave activity flux
\citep[e.g.,][]{Edmon80}. Because meridional wave activity fluxes in
the upper troposphere can be expected to diverge poleward of $\phi_e$
(where vertical wave activity fluxes converge) and because the
meridional wave activity flux divergence is proportional to the eddy
momentum flux convergence, there is upper-tropospheric eddy momentum
flux convergence poleward of $\phi_e$ and divergence equatorward of
$\phi_e$ \citep[e.g.,][]{Held75,Held00b,Simmons78,Edmon80}. At the
latitude $\phi_e$, then, the eddy momentum flux divergence
$\mathcal{S}$ in the upper troposphere changes sign. Because the local
Rossby number is generally small near the subtropical termini of the
Hadley circulation, the zonal momentum balance \eqref{e:zon_mom} there
is approximately
\begin{equation}
  f \Bar v \approx \mathcal{S},
\end{equation}
so that a change in sign in $\mathcal{S}$ implies a change in sign in
the meridional mass flux: the latitude $\phi_e$ marks the transition
between the Hadley cells, near whose subtropical termini
$\mathcal{S}>0$ and $\Bar v$ is poleward, and the Ferrel cells, in
which $\mathcal{S} < 0$ and $\Bar v$ is equatorward
(Figs.~\ref{f:hadley_seasonal} and \ref{f:gcm_hadley}).

With this notion of what terminates the Hadley circulation, it remains
to relate the height reached by substantial eddy entropy fluxes to the
mean temperature structure and other mean fields and parameters. In
dry atmospheres, the supercriticality
\begin{equation}
  S_c = -\frac{f}{\beta}\frac{\partial_y \Bar\theta_s}{\Delta_v} 
  \sim \frac{\Bar p_s - \Bar p_e}{\Bar p_s - \Bar p_t},
  \label{e:sc}
\end{equation}
is a nondimensional measure of the pressure range over which eddy
entropy fluxes extend (\citet{Schneider06a,Schneider07b}; see
\citet{Held78b} for a similar measure in quasigeostrophic
theory). Here, $\beta=2\Omega a^{-1} \cos\phi$ is the meridional
derivative of the Coriolis parameter $f$; $\Bar \theta_s$ is the mean
surface or near-surface potential temperature; $\Delta_v$ is a bulk
stability measure that depends on the near-surface static stability;
and $\Bar p_s$, $\Bar p_t$, and $\Bar p_e$ are the mean pressures at
the surface, at the tropopause, and at the level up to which eddy
entropy fluxes extend. Consistent with the preceding discussion, the
Hadley circulation in dry GCM simulations, in parameter regimes
comparable with Earth's, extends up to the latitude $\phi_e$ at which
the supercriticality \eqref{e:sc}, evaluated locally in latitude,
first exceeds a critical $O(1)$ value \citep{Korty08}. In particular,
the Hadley circulation generally widens as the bulk stability
$\Delta_v$ increases, consistent with the increase in $\Delta_v$ at
the subtropical termini being primarily compensated by an increase in
$f/\beta = a \tan \phi_e$.

There are two challenges to obtain a closed theory of the width of the
Hadley circulation from these results. First, for dry atmospheres, the
mean fields in the supercriticality \eqref{e:sc} need to be related to
the mean meridional circulation and eddy fluxes, which determine them
in concert with radiative processes. For a Hadley circulation whose
upper branches approach the angular momentum--conserving limit, an
expression for the width can be derived in which the meridional
surface potential temperature gradient no longer appears explicitly
\citep{Held00b}; however, because the Hadley circulation generally
does not approach the angular momentum--conserving limit, the
resulting expression does not accurately account for changes in the
width, even in dry GCM simulations
\citep{Walker06,Schneider06b,Korty08}. (Some recent papers have
advocated similar expressions to account for the relatively modest
changes in the Hadley circulation width seen in simulations of climate
change scenarios \citep[e.g.,][]{Lu07,Frierson07d}, but the results
from the much broader range of dry GCM simulations in \citet{Walker06}
and \citet{Korty08} show that these expressions cannot be generally
adequate.) In addition to the meridional surface potential temperature
gradient, one needs to close for the near-surface static stability,
which likewise depends on the flow. The static stability at the
subtropical termini of the Hadley circulation cannot simply be viewed
as controlled by convection, as in the deep tropics, but it is
influenced by the mean meridional circulation and eddy fluxes.

Second, for moist atmospheres, the supercriticality \eqref{e:sc} does
not generally give a good estimate of the height reached by
substantial eddy entropy fluxes \citep{Schneider08c}. The difficulties
in relating the static stability at the subtropical termini of the
Hadley circulation to mean flows and eddy fluxes are exacerbated in
moist atmospheres, in which it is unclear what the effective static
stability is that eddy fluxes experience, how that effective static
stability is controlled, and how it relates to the depth of eddy
entropy fluxes. We currently do not have theories of the static
stability and Hadley circulation width that are adequate for moist
atmospheres.\footnote{In addition to the mechanisms sketched here, the
  width of the Hadley circulation may also change in response to
  changes in upper-tropospheric wave dynamics that may be caused by
  lower-stratospheric changes associated with ozone depletion or
  increased concentrations of greenhouse gases \citep{Chen07b}. See
  section~\ref{s:st_tracks}.}

\begin{figure}[!tbh]
  \noindent\centerline{\includegraphics{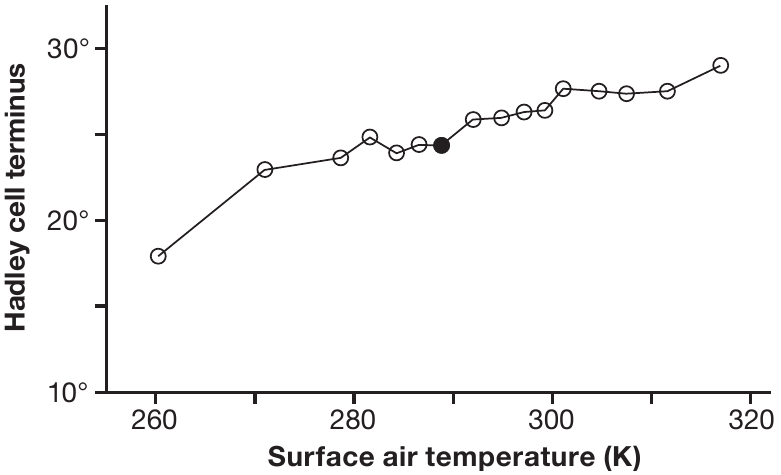}}
  \caption{Hadley circulation width vs global-mean surface temperature
    in idealized GCM simulations. Shown is the latitude of the
    subtropical terminus of the Hadley circulation, defined as the
    latitude at which the mass flux streamfunction at approximately
    $725\,\mathrm{hPa}$ is zero. The termination latitudes in both
    hemispheres are averaged.}
  \label{f:hc_width}
\end{figure}
In the idealized GCM simulations presented throughout this paper, the
width of the Hadley circulation increases modestly with surface
temperature. The Hadley circulation extends to $18^\circ$ latitude in
the coldest simulation, to $24^\circ$ in the reference simulation, and
to $29^\circ$ latitude in the warmest simulation. The Hadley
circulation in the reference simulation is narrower than Earth's---at
least in part because ocean heat transport is neglected, so that
meridional surface temperature gradients in the tropics are larger
than on Earth. The increase in the width of the Hadley circulation
with surface temperature is qualitatively consistent with the notion
that baroclinic eddy fluxes terminate the Hadley circulation, and that
the latitude at which they reach the upper troposphere moves poleward
as the subtropical static stability increases, in part but not
exclusively because the moist-adiabatic lapse rate decreases with
temperature. However, the increase in the width is not quantitatively
consistent with the arguments for dry atmospheres. Devising a theory
that accounts for these results remains as one of the fundamental
challenges in completing a theory of the general circulation of moist
atmospheres.

\section{Extratropical circulations}\label{s:extratropics}

One measure of the importance of water vapor and latent heat release
in extratropical circulations is the fraction of the poleward energy
flux that takes the form of a latent heat flux. In the present
climate, this is about half of the total atmospheric energy flux in
midlatitudes \citep{Pierrehumbert02,Trenberth03a}, indicating a
significant role for water vapor in extratropical dynamics.  But
whereas water vapor plays an unambiguously important role in tropical
dynamics, its role in extratropical dynamics is less clear. 

Moist convection in the extratropics is not as ubiquitous as it is in
the tropics (over oceans, it primarily occurs in fronts of large-scale
eddies), so that the precise dynamical role of water vapor in the
extratropics is unclear. The importance of water vapor in
extratropical dynamics may depend strongly on the warmth of the
climate considered, as surface temperatures in the extratropics
respond more strongly to climate changes than in the tropics, and the
saturation vapor pressure and thus the near-surface specific humidity
depend nonlinearly on temperature. Water vapor likely has a much
reduced dynamical role in the extratropics of cold climates, such as
that of the LGM, and a correspondingly greater role in hothouse
climates.  The unclear role of water vapor in extratropical dynamics
in the present climate and its changed importance in colder or warmer
climates are principal challenges in understanding extratropical
circulations and their response to climate changes.

\subsection{Transient eddy kinetic energy}\label{s:eke}

Several lines of evidence point to an influence of latent heat release
on the structure and amplitude of extratropical storms, ranging from
studies of individual cyclones \cite[e.g.,][]{Reed88, Wernli02}, to
theoretical considerations of the effect of water vapor on the mean
available potential energy \citep{Lorenz78}.  The mean available
potential energy is a measure of the energy available to midlatitude
transient eddies through adiabatic air mass rearrangements
\citep[][chapter~14]{Peixoto92}. It is always greater when the
potential release of latent heat in condensation of water vapor is
taken into account. For a zonal-mean state similar to that of the
present climate, the mean moist available potential energy is roughly
30\% greater than the mean dry available potential energy
\citep{Lorenz79}.  Latent heat release also increases the linear
growth rate of moist baroclinic eddies \citep{Bannon86, Emanuel87b},
leads to greater peak kinetic energy in lifecycle studies of
baroclinic eddies \citep{Gutowski92}, and contributes positively to
the budget of eddy available potential energy in Earth's storm tracks
\citep{Chang02}.

\begin{figure}[!tbh]
  \noindent\centerline{\includegraphics{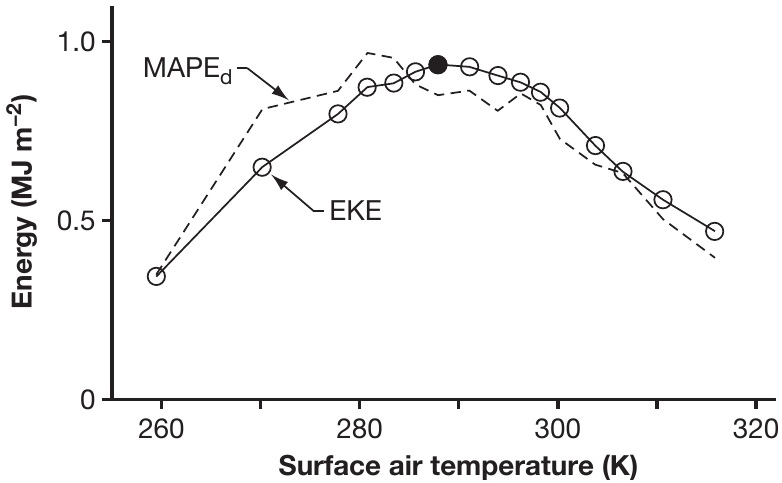}}
  \caption{Total eddy kinetic energy (EKE) (solid line with circles),
    and rescaled dry mean available potential energy
    ($\mathrm{MAPE_d}$) (dashed line) vs global-mean surface
    temperature in idealized GCM simulations.  Averages are taken over
    baroclinic zones in both hemispheres.  $\mathrm{MAPE_d}$ is
    evaluated for the troposphere and is rescaled by a constant factor
    of $2.4$.  (Adapted from \citet{OGorman08c}.)}
  \label{f:eke}
\end{figure}
It is therefore somewhat surprising that the total (vertically
integrated) eddy kinetic energy scales approximately linearly with the
\emph{dry} mean available potential energy in the idealized GCM
simulations (Fig.~\ref{f:eke}).  The energies shown are averaged
meridionally over baroclinic zones, which are here taken to be
centered on maxima of the eddy potential temperature flux and to have
constant width $L_Z$ corresponding to $30^\circ$ latitude
\citep{OGorman08c}.  Both the eddy kinetic energy and the dry mean
available potential energy have a maximum for a climate close to that
of present-day Earth and are smaller in much warmer and much colder
climates (Fig.~\ref{f:eke}).  Similar behavior is found for the
near-surface eddy kinetic energy: surface storminess likewise is
maximal in a climate close to that of present-day Earth
(Fig.~\ref{f:eke_vs_lat}).  Broadly consistent with the idealized GCM
simulations, simulations with comprehensive GCMs suggest that
extratropical storms change only modestly in strength when the present
climate changes \citep{Geng03, Yin05, Bengtsson06, Bengtsson09}, and
they can be weaker both in glacial climates \citep{Li08} and in warm,
equable climates \citep[e.g.,][]{Rind86,Korty07a}.

To understand why the energies in the idealized GCM are maximal in a
certain climate, it is useful to consider the approximate dry mean
available potential energy 
\begin{equation}\label{e:mape} 
  \mathrm{MAPE_d} \approx \frac{c_p}{24 \,g} \,
  \Delta p_t \, L_Z^2 \,  \Gamma \, (\partial_y \Bar\theta)^2, 
\end{equation}
obtained as a scaling approximation of Lorenz's \citeyearpar{Lorenz55}
mean available potential energy \citep{Schneider81, Schneider08b,
  OGorman08c}.  Here, $\Delta p_t=\Bar p_s - \Bar p_t$ is the pressure
depth of the troposphere,
\begin{equation}\label{e:static_stab}
  \Gamma = -\frac{\kappa}{p} (\partial_p \Bar\theta)^{-1}
\end{equation} 
is an inverse measure of the dry static stability, $\partial_y
\Bar\theta$ is the mean meridional potential temperature gradient,
$\kappa$ is the adiabatic exponent, and $g$ is the gravitational
acceleration.  The meridional potential temperature gradient and
inverse static stability are understood to be averaged vertically over
the depth of the troposphere and meridionally over baroclinic zones,
in addition to zonally and temporally \citep{OGorman08c}.  According
to the approximation \eqref{e:mape}, $\mathrm{MAPE_d}$ increases with
increasing meridional potential temperature gradients and tropopause
height and with decreasing static stability.  In the idealized GCM
simulations, several factors conspire to lead to the non-monotonic
behavior of $\mathrm{MAPE_d}$:
\begin{enumerate}
\item[(i)] As the climate warms relative to the reference climate, the
  vertically averaged meridional potential temperature gradient
  decreases and the static stability increases (see
  Fig.~\ref{f:static_stab} below). These changes in the thermal
  structure of the troposphere primarily result from increased
  poleward and upward transport of latent heat. 

  There is also a countervailing increase in tropopause height (it
  changes for the reasons discussed in section~\ref{s:tropopause}),
  but the combined changes in static stability and temperature
  gradient are larger and result in a decrease in $\mathrm{MAPE_d}$.
\item[(ii)] As the climate cools relative to the reference climate,
  the near-surface meridional potential temperature gradient increases
  strongly. The vertically averaged meridional potential temperature
  gradient also increases, albeit less strongly than the near-surface
  gradient because the tropical temperature lapse rate, which is
  approximately moist adiabatic, increases, whereas the extratropical
  lapse rate, which is at least partially determined by baroclinic
  eddies (see section~\ref{s:stratification} below), decreases. In
  $\mathrm{MAPE_d}$, the increase in the vertically averaged
  meridional potential temperature gradient is overcompensated by
  decreases in the tropopause height 
  and by the increase in the
  extratropical static stability.
\end{enumerate}

It is noteworthy that changes in the eddy kinetic energy need not be
of the same sign as changes in the near-surface meridional temperature
gradient, contrary to what is sometimes assumed in discussions of
extratropical storminess (e.g., at the LGM). In the idealized GCM
simulations, the near-surface meridional temperature gradient
decreases monotonically as the climate warms, whereas the eddy kinetic
energy (and $\mathrm{MAPE_d}$) change non-monotonically.

The scaling of the eddy kinetic energy with the dry mean available
potential energy intimates that water vapor dynamics affects the eddy
kinetic energy in the idealized GCM primarily through its effect on
the thermal structure of the troposphere, rather than through direct
effects of latent heat release on eddies. Because extratropical water
vapor dynamics generally decreases meridional potential temperature
gradients and increases the (dry) static stability, it primarily damps
eddies, rather than energizing them, as one might have inferred from
the fact that in Earth's storm tracks, latent heat release contributes
positively to the budget of eddy available potential energy
\citep[cf.][]{Chang02}. Although it may seem surprising and is largely
an empirical result that the eddy kinetic energy scales with the dry
mean available potential energy and thus depends on the dry static
stability, there are several plausible reasons for this
\citep{OGorman08c}. For example, the 30\% difference between mean dry
and moist available potential energies that \citet{Lorenz79} found for
the present climate largely arises owing to water vapor in tropical
low-level regions, which may not be important for midlatitude
eddies. Additionally, changes in the effective moist static stability
that midlatitude eddies experience may generally scale with changes in
the dry static stability if the effective moist static stability is a
weighted average of a dry stability and a smaller moist stability in
updrafts \citep{Emanuel87b}, and if the weighting coefficients (e.g.,
the area fractions of updrafts and downdrafts) do not change
substantially with climate.

Does the eddy kinetic energy always scale with the dry mean available
potential energy, as in the idealized GCM, or can latent heat release
directly energize the statistically steady state of baroclinic eddies?
\citet{Lapeyre04} analyzed the moist eddy available potential energy
budget of a two-layer quasigeostrophic model with water vapor in the
lower layer. In the model, increases in the production of moist eddy
available potential energy associated with latent heat release are
primarily balanced by water vapor diffusion and dehumidification
processes, rather than by conversion to eddy kinetic energy, implying
an inefficient heat engine.  For very strong latent heat release, a
vortex-dominated regime emerged that had no analog in a corresponding
dry model.  While the study of \citet{Lapeyre04} provides some
guidance to the possible role of water vapor in the dynamics of
baroclinic eddies in a statistically steady state, it is difficult to
relate these results to the behavior of moist baroclinic eddies in
general circulation models or in the real atmosphere.

We have used averages of the eddy kinetic energy to give a general
description of the effect of water vapor on the amplitude of
baroclinic eddies.  However, this does not tell us about the possible
effects of changes in latent heat release, for example, on mesoscale
wind extremes or on the local energy of cyclones in zonally confined
storm tracks. Changes in the structure of baroclinic eddies due to
latent heat release also affect the magnitude and extent of updrafts
\citep{Emanuel87b, Zurita-Gotor05}, which can be expected to influence
extratropical precipitation and its extremes. Extratropical mean
precipitation and precipitation extremes generally increase in
intensity as the climate warms, albeit at a smaller rate than the mean
specific humidity \citep{OGorman09a,OGorman09b}.

\subsection{Position of  storm tracks}\label{s:st_tracks}

\begin{figure}[!tbh]
  \noindent\centerline{\includegraphics{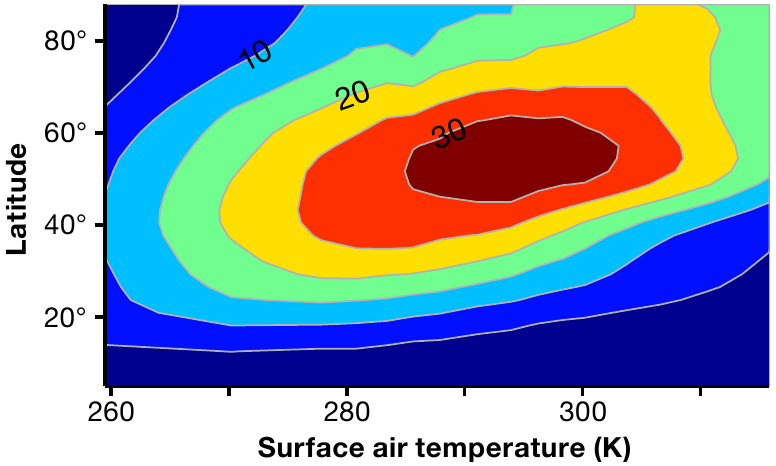}}
  \caption{Near-surface eddy kinetic energy (contour interval
    $5\,\mathrm{kJ \, m^{-2}}$) as a function of global-mean surface
    temperature and latitude in idealized GCM simulations.  The
    near-surface eddy kinetic energy is integrated from the surface to
    $\sigma=0.9$.}
  \label{f:eke_vs_lat}
\end{figure}
The extratropical storm tracks generally shift poleward as the climate
warms in simulations of climate change scenarios \citep{Yin05,
  Bengtsson06}.  They also shift poleward as the climate warms in the
idealized GCM simulations \citep{OGorman08b}, provided storm tracks
are identified with regions of large near-surface eddy kinetic energy
(Fig.~\ref{f:eke_vs_lat}).\footnote{The changes in eddy kinetic energy
  at upper levels are complicated by changes in jet structure, and the
  mean near-surface westerlies actually shift equatorward as the
  climate warms over part of the range of simulations.  This suggests
  that the changes in eddy-mean flow interaction in the simulations
  are not straightforward and deserve further investigation.} 

Attempts have been made to relate changes in the position of
extratropical storm tracks to changes in local measures of baroclinic
instability. The Eady growth rate is typically used as the measure of
baroclinic instability \citep[e.g.,][]{Lindzen80a, Hoskins90, Geng03,
  Yin05, Li08, Brayshaw08}. It depends on the meridional potential
temperature gradient and the dry static stability and is similar to
the square root of the mean available potential energy \eqref{e:mape}.
\citet{Yin05} found that changes in the Eady growth rate in climate
change simulations seemed to account for a poleward shift in the eddy
kinetic energy maximum, and that more of the change was related to the
meridional potential temperature gradient than to the static
stability.  However, it is not clear how the local linear growth rate
of baroclinic instability relates to the distribution of eddy kinetic
energy in a statistically steady state. For example, the Eady growth
rate in the idealized GCM simulations typically has two maxima as a
function of latitude, one near the subtropical terminus of the Hadley
cell, and one in midlatitudes. In warm simulations, the subtropical
maximum in growth rate is the hemispheric maximum, but it is located
equatorward of the storm track.

Latent heat release helps to set the mean thermal structure of the
troposphere and thus indirectly affects the dry Eady growth rate and
other measures of baroclinicity.  But it can also directly affect the
growth rate of baroclinic instability, an effect which probably must
be taken into account when considering climate changes, given how
rapidly precipitable water increases with temperature.
\citet{Orlanski98} proposed using an approximate result for the moist
baroclinic instability growth rate based on the work of
\citet{Emanuel87b} but he found that the inclusion of latent heat
release only modestly affects the growth rates in the winter storm
track. It remains unclear how growth rates of baroclinic instability
depend on the mean state of a moist atmosphere, and how they relate to
storm track position in other seasons or in very warm climates.

\citet{Chen07b} proposed a different approach to understanding shifts
in the storm tracks, based on considering changes in the momentum
fluxes associated with upper-tropospheric eddies.  Key to the
mechanism they propose are changes in upper-tropospheric and
lower-stratospheric zonal winds that are linked, by thermal wind
balance, to changes in the thermal structure near the tropopause. For
example, increases in the concentration of greenhouse gases generally
lead to lower-stratospheric cooling and upper-tropospheric warming,
which imply a strengthening of lower-stratospheric zonal (westerly)
winds around the poleward and downward sloping extratropical
tropopause. Such changes can modulate the phase speed of
upper-tropospheric eddies and may, via a shift in their critical
latitude, lead to a shift in the position of storm tracks. Unlike the
other mechanisms we discussed, this mechanism relies on radiative
changes in the lower stratosphere, which are not well represented by
the simplified radiation scheme of our idealized GCM. Since the
dynamics of upper-tropospheric eddies are largely unaffected by latent
heat release, the mechanism also does not allow for a direct role for
water vapor dynamics.

There currently is no comprehensive theory for the position of storm
tracks, even in the zonal mean. It is even less clear what determines
the longitudinal extent of zonally varying storm tracks
\citep{Chang02}. 

\subsection{Poleward energy flux}

\begin{figure}[!tbh]
  \noindent\centerline{\includegraphics{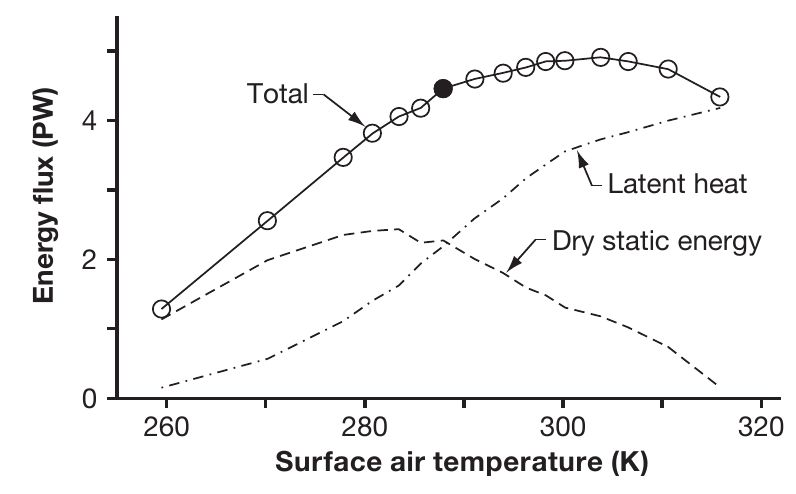}}
  \caption{Vertically integrated poleward energy flux (solid line and
    circles) at $50^\circ$ latitude vs global-mean surface temperature
    and decomposition into dry static energy flux (dashed line) and
    latent heat flux (dash-dotted line).  (Adapted from
    \citet{OGorman08b}.)}
  \label{f:poleward_flux}
\end{figure}

The poleward energy flux in the extratropics, effected primarily by
eddies, is essential to the maintenance of climate, particularly in
high latitudes. The total energy flux can be divided into the
atmospheric fluxes of dry static energy, $c_p T + g z$, and latent
heat, $L q$, plus the ocean heat flux; the kinetic energy flux is
negligible in both the atmosphere and oceans \citep{Peixoto92}.  In
the idealized GCM simulations, the relative contributions to the
extratropical poleward energy flux from dry static energy and latent
heat vary strongly with climate (Fig.~\ref{f:poleward_flux}). The dry
static energy flux dominates in cold climates; the latent heat flux
dominates in warm climates. The total poleward energy flux does not
remain constant as the climate varies, but it increases from the
coldest to moderately warm simulations and decreases again in the
warmest simulations. Close to the reference simulation, there is some
compensation between opposing changes in latent heat and dry static
energy fluxes, but the compensation is not exact and not a general
feature of climate changes: in cold simulations, for example, changes
in latent heat and dry static energy fluxes have the same sign.  This
stands in contrast to the almost exact compensation between changes in
poleward energy flux components that \citet{Frierson07a} found in a
similar idealized GCM as they varied the amount of water vapor in the
atmosphere, keeping radiative transfer parameters fixed. The
difference in behavior most likely results from the difference in how
the climate is varied (changing longwave optical thickness vs changing
water vapor concentrations while keeping radiative parameters
fixed). Figure~\ref{f:poleward_flux} shows that a compensation between
changes in poleward energy flux components cannot generally be
expected in response to climate changes such as those induced by
changes in greenhouse gas concentrations.

How the poleward energy flux changes with climate is relevant to
several fundamental questions, including the question of how small the
pole-equator temperature contrast can get in equable climates,
given the insolation distribution. For very warm climates, it becomes
essential to understand the scaling of the poleward latent heat flux
because it dominates the total poleward energy flux in such climates.
In the extratropics, the poleward latent heat flux is dominated by the
eddy component $\mathcal{F}_e = \{L \overline{v'q'}\}$, which scales
like
\begin{equation}\label{e:lh_flux}
  \mathcal{F}_e 
  \sim   L v_e \, q_{\mathrm{ref}}\, p_0/g, 
\end{equation}
where $q_{\mathrm{ref}}$ is a subtropical reference specific humidity,
$v_e$ is an eddy velocity scale, and $p_0$ is the mean surface
pressure \citep{Pierrehumbert02, Caballero05, OGorman08b}.  The
scaling derives from the assumption that the eddy flux of latent heat
is effected by eddies that pick up water vapor in or near the boundary
layer in the subtropics and transport it poleward and upward along
approximately isentropic paths, along which air masses cool, the
specific humidity reaches saturation, and the water vapor condenses
out. According to the scaling \eqref{e:lh_flux}, the decrease in eddy
kinetic energy (and thus in $v_e$) in warm climates (Fig.~\ref{f:eke})
plays a critical role in limiting the poleward latent heat flux and
hence the minimum attainable pole-equator temperature contrast
\citep{Caballero05}.  Since the eddy kinetic energy itself depends on
the pole-equator temperature contrast and on the static stability (as
discussed in section~\ref{s:eke}), and since these in turn depend on
water vapor dynamics, interesting dynamical feedbacks in which water
vapor plays a major role are conceivable.  The scaling
\eqref{e:lh_flux} generally accounts well for the eddy latent heat
flux in the idealized GCM simulations, except in the warmest
simulations, in which it overestimates the latent heat flux
\citep{OGorman08b}. More sophisticated scalings may be needed for the
poleward latent heat flux in very warm climates or at high latitudes;
analyses of how water vapor is transported along isentropes and
condenses may be useful in this regard
\citep{Pierrehumbert07,OGorman06}.

The poleward dry static energy flux in the idealized GCM simulations
changes non-monotonically with global-mean surface temperature
(Fig.~\ref{f:poleward_flux}). As for the eddy kinetic energy, changes
in the dry static energy flux can have the opposite sign of changes in
the near-surface meridional temperature gradient, contrary to what is
sometimes assumed. In the idealized GCM simulations, the dry static
energy flux is maximal in climates slightly colder than that of
present-day Earth, as is the eddy kinetic energy and the dry mean
available potential energy (cf.\ Fig.~\ref{f:eke}). Indeed, in dry
atmospheres in which baroclinic eddies modify the thermal
stratification, the eddy flux of dry static energy, which dominates
the extratropical dry static energy flux, scales with
$\mathrm{MAPE_d}/\Gamma^{1/2}$, that is, with the mean available
potential energy modulated by a weak dependence on the static
stability $\Gamma^{-1}$ \citep{Schneider08b}. This scaling derives
from assuming that the eddy kinetic energy scales with
$\mathrm{MAPE_d}$, that eddy kinetic energy and eddy available
potential energy are equipartitioned, and that the eddy flux of dry
static energy can be related to the eddy kinetic energy and eddy
available potential energy.  These assumptions are sufficiently well
satisfied in the idealized GCM simulations that the scaling correctly
suggests a climatic maximum in the dry static energy flux.

\subsection{Thermal stratification}\label{s:stratification}

\begin{figure}[!tbh]
  \noindent\centerline{\includegraphics{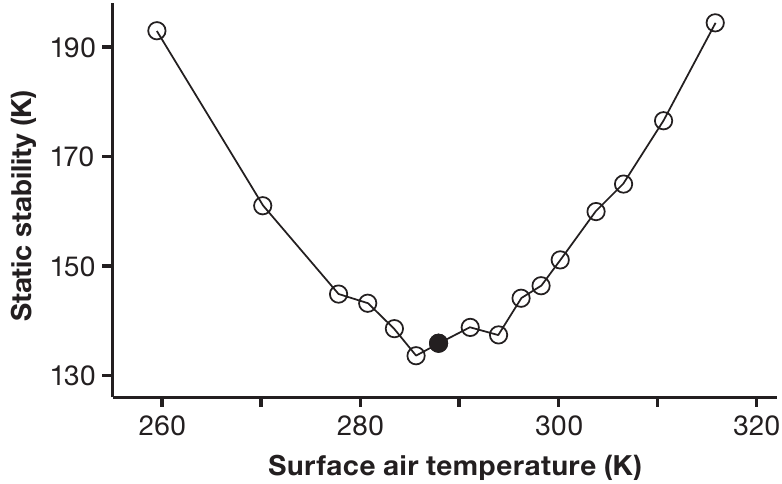}}
  \caption{Extratropical dry static stability vs global-mean surface
    temperature in idealized GCM simulations. The static stability is
    computed by averaging $\Gamma^{-1}$ (Eq.~\ref{e:static_stab})
    vertically over the troposphere and meridionally over baroclinic
    zones in both hemispheres.  (Adapted from \citet{OGorman08c}.)}
  \label{f:static_stab}
\end{figure}

The mean thermal stratification of the extratropical troposphere
influences important climatic features such as the eddy kinetic
energy, the position of storm tracks, and the poleward energy flux.
Water vapor dynamics affects the thermal stratification through latent
heat release in moist convection and in large-scale condensation. In
the idealized GCM simulations, the extratropical static stability
increases as the climate warms relative to the reference climate,
largely because the poleward and upward transport of latent heat
strengthens as the climate warms. However, the extratropical static
stability also increases as the climate cools relative to the
reference climate (Fig.~\ref{f:static_stab}). 

In colder climates, the amount of water vapor in the extratropical
atmosphere is small, and a dry theory for the thermal stratification
accounts for the simulation results: The static stability is
proportional to the meridional surface temperature gradient multiplied
by $f/\beta$ evaluated at the storm track latitude, such that the
supercriticality \eqref{e:sc} averaged over the extratropics satisfies
$S_c \sim 1$ \citep{Schneider06a,Schneider08c}. The static stability
thus decreases as the meridional surface temperature gradient
decreases and the climate warms ($f/\beta$ at the storm track latitude
increases slightly as the storm tracks shift poleward, but this is
overcompensated by the surface temperature gradient
decrease). However, latent heat release plays an increasingly
important role in the maintenance of the extratropical stratification
as the climate warms, primarily through large-scale latent heat
fluxes. In warmer climates, the dry theory with $S_c \sim 1$ is no
longer applicable.  Static stabilities are generally greater in these
warm and moist climates than the dry theory would predict
\citep{Schneider08c}.

In addition to large-scale latent heat fluxes, moist convection
becomes more prevalent in the extratropics in warmer climates. A
theory that posits a central role for baroclinic eddies and moist
convection has been proposed for the extratropical thermal
stratification \citep{Juckes00, Frierson06a, Frierson08a}. However, a
direct role for extratropical moist convection in setting the thermal
stratification in the idealized GCM simulations is ruled out by a set
of simulations in which the temperature lapse rate toward which the
convection scheme relaxes temperature profiles was artificially
rescaled to be lower (more stable) than the moist-adiabatic lapse
rate; instead of moist convection, large-scale latent heat fluxes
appear to be crucial for the extratropical thermal stratification
\citep{Schneider08c}. Nonetheless, these results from idealized
aquaplanet simulations do not preclude an important role for
extratropical (possibly slantwise) moist convection in setting the
thermal stratification seasonally or regionally in Earth's atmosphere
\citep{Emanuel88a}. For example, moist convection does appear to
control the extratropical thermal stratification over some land
surfaces in summer \citep{Korty07}.  The formulation of a general
theory of the extratropical thermal stratification that accounts for
latent heat release, in moist convection and in large-scale fluxes,
remains an outstanding challenge.

Through its effect on the thermal structure of the troposphere and the
poleward energy flux, together with indirect (and possibly direct)
effects on the extratropical storm tracks, water vapor dynamics play
an important role in extratropical circulations, except in very cold
climates.  To make further progress understanding how extratropical
atmospheric dynamics change with climate, it will be necessary to
develop theories for extratropical dynamics that take direct account
of latent heat release. Such theories must be reducible to existing
theories for dry dynamics, but it is unclear to what extent they can
be developed through generalization of concepts from dry dynamics
(e.g., replacing dry static stabilities by effective moist static
stabilities).

\section{Summary and open questions}\label{s:summary}

We have presented an overview of dynamic effects of water vapor in the
global circulation of the atmosphere and in climate changes,
illustrated by simulations of a broad range of climates with an
idealized GCM. With a review of global energetic constraints on
hydrologic variables as point of departure, we discussed how water
vapor dynamics affects the tropical gross upward mass flux, how the
Hadley circulation changes with climate, and how aspects of
extratropical circulations, such as extratropical storminess and the
poleward energy transport, relate to and influence the mean climate
state. Central conclusions were:
\begin{enumerate}
\item Changes in global-mean evaporation and precipitation and in
  near-surface relative humidity are strongly energetically
  constrained. Near the present climate, global-mean evaporation and
  precipitation can increase with surface temperature at a rate of
  $O(2\%\,\mathrm{K^{-1}})$, and the near-surface relative humidity
  can change by $O(1\%\,\mathrm{K^{-1}})$.
\item Because changes in near-surface relative humidity are small and
  most water vapor is concentrated near the surface, precipitable
  water increases with surface temperature approximately at the
  Clausius-Clapeyron rate at which saturation specific humidity
  increases. Near the present climate, this rate is
  $6$--$7\%\,\mathrm{K^{-1}}$.
\item Although the water vapor cycling rate generally decreases as the
  climate warms, except in very cold climates, the tropical gross
  upward mass flux does not necessarily decrease at a similar rate, or
  at all. Rather, the tropical gross upward mass flux may depend on
  precipitation and the moist-adiabatic static stability of the
  tropical atmosphere, which changes more slowly with temperature than
  precipitable water.
\item The Hadley circulation generally widens and increases in height
  as the climate warms. Changes in its strength are more complex. They
  are constrained by the zonal momentum balance and the strength of
  eddy momentum fluxes. Near the present climate, the Hadley cell
  likely weakens as the climate warms; however, it may also weaken as
  the climate cools, in part because the eddy momentum fluxes, whose
  strength is related to the extratropical eddy kinetic energy, can
  change non-monotonically with climate.
\item The extratropical transient eddy kinetic energy, a measure of
  storminess, scales with the dry mean available potential
  energy. Near the present climate, both energies decrease as the
  climate warms, because meridional potential temperature gradients
  decrease and the static stability increases as the poleward and
  upward transport of latent heat strengthens. In colder climates,
  however, both energies can also decrease as the climate cools.
\item Storm tracks generally shift poleward as the climate warms. 
\item The poleward latent heat flux in the extratropics generally
  increases as the climate warms, but the dry static energy flux can
  change non-monotonically. The total poleward energy flux, the sum of
  the two, can also change non-monotonically, suggesting there may
  exist a limit on how small pole-equator temperature contrasts can
  become in equable climates.
\item The behavior of the extratropical static stability is
  complex. Strengthening poleward and upward latent heat transport in
  warmer and moister climates can increase the static stability. And
  strengthening meridional surface temperature gradients in colder and
  drier climates can also lead to an increase in static stability.
\end{enumerate}
A recurring theme was that although hydrologic variables such as
global-mean precipitable water and precipitation change monotonically
with surface temperature, dynamical variables such as the tropical
gross upward mass flux or the extratropical eddy kinetic energy need
not change monotonically; they can be weaker than they presently are
both in much warmer and in much colder climates.

A number of questions have remained open, chief among them:
\begin{enumerate}
\item How do changes in the mean meridional circulation and in eddy
  momentum fluxes interact to control how the strength of the Hadley
  circulation changes with climate?
\item How does the width of the Hadley circulation depend on mean
  fields such as meridional temperature gradients, the specific
  humidity, and the (subtropical) static stability?
\item Can latent heat release directly energize the statistically
  steady state of extratropical eddies? Or is its main effect through
  modifications of the mean state of the atmosphere?
\item What controls the position of storm tracks and their poleward
  shift as the climate warms? More generally, how do eddy kinetic
  energies and other eddy fields depend on mean fields, and what
  controls their variations with latitude?
\item What controls the static stability of the subtropical and
  extratropical atmosphere?
\end{enumerate}
The lack of a theory for the subtropical and extratropical static
stability runs through several of the open questions. Devising a
theory that is general enough to be applicable to relatively dry and
moist atmospheres remains as one of the central challenges in
understanding the global circulation of the atmosphere and climate
changes. 


%
%
%
%
%
%

%
%
%
%

\begin{acknowledgments}
  We are grateful for support by the Davidow Discovery Fund, the
  National Science Foundation (Grant ATM-0450059), and a David and
  Lucile Packard Fellowship. The simulations shown were performed on
  Caltech's Division of Geological and Planetary Sciences Dell
  cluster. Portions of section~\ref{s:constraints} appeared previously
  in the Proceedings of the 15th 'Aha Huliko'a Hawaiian Winter
  Workshop \citep{Schneider07y}. We thank Ian Eisenman, Yohai Kaspi,
  Tim Merlis, Steven Sherwood, and two reviewers for helpful comments
  on a draft of this paper.
\end{acknowledgments}

%
%
%
%
%
%
%
%
%
%



%
%

\end{article}




%
%
%
%
%
%


\end{document}